\documentclass[11pt,preprint,floatfix]{revtex4}
\usepackage{amssymb}

\usepackage[dvips]{graphicx}
\usepackage{subfigure}
\usepackage{morefloats}

\input{psfig.sty}

\begin{document}

\title{Scaling in Non-stationary time series I}
\author{M. Ignaccolo$^{1}$\footnote{Corresponding Author.\newline \textit{Mailing Address}: Center for Nonlinear Science, University of North Texas,
P.O. Box 311427, Denton, Texas, 76203-1427 . \newline \textit{Phone}: +1 940 565 3280 . \newline \textit{E-mail Address}: stellina99@earthlink.net . }, P. Allegrini$^{2}$, P. Grigolini$^{1,3,4}$, P. Hamilton$^{5}$, B. J. West$^{6}$}
\address{$^{1}$Center for Nonlinear Science, University of North Texas,
P.O. Box 311427, Denton, Texas, 76203-1427} 
\address{$^{2}$ Istituto di Linguistica Computazionale del Consiglio Nazionale delle
Ricerche,\\
 Area della Ricerca di Pisa-S. Cataldo, Via Moruzzi 1, 
56124, Ghezzano-Pisa, Italy } 
\address{$^{3}$Dipartimento di Fisica dell'Universit\`{a} di Pisa and INFM 
Via Buonarroti 2, 56127 Pisa, Italy } 
\address{$^{4}$Istituto di Biofisica del Consiglio Nazionale delle
Ricerche,\\ Area della Ricerca di Pisa-S. Cataldo, Via Moruzzi 1,
56124, Ghezzano-Pisa, Italy } 
\address{$^{5}$ Center for Nonlinear Science, Texas Woman's University, 
P.O. Box 425498, Denton, Texas 76204} 
\address{$^{6}$ Physics Department,
Duke University, P.O. Box 90291, Durham, North Carolina 27708
and US Army Research Office, 
Mathematics Division, 
Research Triangle Park, NC 27709}
\date{\today}

\begin{abstract}
Most data processing techniques, applied to biomedical and sociological time
series, are only valid for random fluctuations that are stationary in time.
Unfortunately, these data are often non stationary and the use of techniques
of analysis resting on the stationary assumption can produce a wrong
information on the scaling, and so on the complexity of the process under
study. Herein, we test and compare two techniques for removing the non-stationary influences from computer generated time series, consisting of the
superposition of a slow signal and a random fluctuation. The former is based
on the method of wavelet decomposition, and the latter is a proposal of this
paper, denoted by us as step detrending technique. We focus our attention on
two cases, when the slow signal is a periodic function mimicking the
influence of seasons, and when it is an aperiodic signal mimicking the
influence of a population change (increase or decrease). For the purpose of
computational simplicity the random fluctuation is taken to be uncorrelated.
However, the detrending techniques here illustrated work also in the case
when the random component is correlated. This expectation is fully confirmed
by the sociological applications made in the companion paper. We also
illustrate a new procedure to assess the existence of a genuine scaling,
based on the adoption of diffusion entropy, multiscaling analysis and the
direct assessment of scaling. Using artificial sequences, we show that the
joint use of all these techniques yield the detection of the real scaling,
and that this is independent of the technique used to detrend the original
signal.
\newline
\newline
\textit{PACS}: 05.45.Tp; 05.40.-a; 87.23.Ge \newline
\textit{keywords}: scaling, multiscaling, diffusion entropy, non-stationary time series, detrending methods.
\end{abstract}

\maketitle

\DeclareGraphicsExtensions{.eps, .ps}

\section{Introduction}\label{intro}

Time series analysis is the backdrop against which most theoretical models
are developed in the biomedical and social sciences. The traditional
assumption made in the engineering literature, and subsequently adopted in
the biophysical, biological and social sciences, is that the
discrete time series variable $\xi\left( t_{j}\right) $ consists of a slowly
varying part $S\left( t_{j}\right) $ and a randomly fluctuating part $\zeta
\left( t_{j}\right) :$

\begin{equation}
\xi\left( t_{j}\right) =S\left( t_{j}\right) +\zeta \left( t_{j}\right) .
\label{intro1}
\end{equation}
The slow, regular variation of the time series is called the \textit{signal}%
, and the rapid erratic fluctuations are called the \textit{noise}. The
implication of this separation of effects is that $S\left( t\right) $
contains information about the system of interest, whereas $\zeta \left(
t\right) $ is a property of the environment and does not contain any
information about the system.



The Science of Complexity is making us acquainted with a different
perspective. Physiological and sociological time series invariably contain
fluctuations, so that when sampled $N$ times the data set $\left\{ \xi
_{j}\right\} $, $j=1,...,N$, appears to be a sequence of random points.
Examples of such data are the interbeat intervals of the human heart \cite
{west2,peng1}, interstride intervals of human gait \cite{hausdorff,west1},
brain wave data from EEGs \cite{west3} and interbreath intervals \cite{szeto}%
, to name a few, and, of course, the Texas teen birth data \cite{damien}
analyzed in the companion paper \cite{companion}. From now on, we shall refer to this paper as paper II. The analysis of the time series in each of these
cases has made use of random walk concepts (see Sec. \ref{tool-diffusion} for the details)
in both the processing of the data and in the interpretation of the results. For example, the
variance of the associated random walk, $\sigma^{2}(t)$, in each of these cases (and many
more) satisfies the property $ \sigma^{2}(t) \propto
t^{2H}$, where $H\neq 1/2$ corresponds to anomalous diffusion. A value of $%
H<1/2$ is interpreted as an antipersistent process in which a step in one
direction is preferentially followed by a reversal of direction. A value of $%
H>1/2$ is interpreted as a persistent process in which a step in one
direction is preferentially followed by another step in the same direction.
A value of $H=1/2$ is interpreted as ordinary diffusion in which the steps
are independent of one another. This would be compatible with the earlier
mentioned concept of environmental noise. In the science of complexity the
fluctuations $\xi _{i}$ are expected to depart from this totally random
condition, since they are expected to have memory and correlation. 
This memory is manifest in inverse power-law spectra of the form $P\left(
\omega \right) \propto 1/\omega ^{\beta +1}$, where the corresponding
correlation function, the inverse transform of the spectrum, is determined
by a Tauberian theorem to be

\begin{equation}
C\left( t_{1},t_{2}\right) \propto \left| t_{1}-t_{2}\right| ^{\beta }.
\label{intro2}
\end{equation}
Here the power-law index is given by $\beta =2H-2.$ Note that the two-point
correlation function depends only on the time difference, thus, the
underlying process is stationary. These properties, as well as a number of
other properties, are discussed for discrete time series by Beran. \cite
{beran}. Another important property of complex systems is that the
associated random walk is not monofractal. For instance the heart beat
variability has been found to be multifractal \cite{ivanov99} as were the
interstride intervals \cite{hausdorff2,griffin}. 
A more accurate way to define the deviation from ordinary diffusion is
ensured by the scaling property. If the time series is long enough,
following the prescription of the pioneer work of Ref. \cite{stanley} we can 
generate a diffusion process out of it. The probability
density function (pdf) of the diffusion process, $p(x,t)$, is expected to
satisfy the scaling condition

\begin{equation}
p(x,t)=\frac{1}{t^{\delta }}F(\frac{x}{t^{\delta }}).
\label{scalingdefinition}
\end{equation}

The deviation from ordinary statistical mechanics, and consequently the
manifestation of complexity, takes place through two distinct quantities.
The first indicator is the scaling parameter $\delta $ departing from the
ordinary value $\delta =0.5,$ which it would have for a simple diffusion
process. The second indicator is the function $F(y)$ departing from the
conventional Gaussian form. The first quantity is usually assessed by
measuring the second moment, or the variance, of the distribution as we did
above. This method of analysis is reasonable only when $F(y)$ maintains its
Gaussian form. If the scaling condition of Eq. (\ref{scalingdefinition})
applies, it is convenient to measure the scaling parameter $\delta $ by the
method of Diffusion Entropy (DE) \cite{de1,de2,de3,de4,de5,de6,de7} that, in
principle, works independently of whether the second moment is finite or
not. The DE method affords many advantages, including that of being totally
independent of a constant bias. Unfortunately, in the presence of a
non-constant bias, the scaling detected by the DE need not indicate the
correlations of the fluctuations, but instead would reflect the influence of
the time-dependent bias.

This important fact, the relationship of the scaling index to the
bias, brings us back to the decomposition of the time series into a signal 
plus noise (see Eq. (\ref{intro1})). We address the theoretical
issues raised by a physical condition where the important information is not
contained in what the engineering literature defines as a signal, but rather
in what is usual defined  as noise. In other words, we adopt
a perspective where the role of the environment might be that of
creating a time-dependent slow component $S(t)$. In this case $S(t)$ could 
be an external forcing, due to seasonal periodicity,
demographic pressure, or to other causes. In any case, we need to
isolate $\xi (t)$ from $S(t)$, to detect the genuine complexity of the
system under study. As we shall show, a process of analysis made
without first establishing this separation might give the
impression of a complexity higher than the real one. This overestimation 
of the scaling index is caused by the non-stationarity of
the time series and inapplicability of the property of Eq. (\ref
{intro2}), implicitly assumed by the method of analysis.

We cannot rule out another important possibility. In complex
phenomena the separation of effects implied by Eq. (\ref{intro1}) may no longer
be appropriate. The low-frequency, slowly changing, part of the spectrum may
be coupled to, and exchange energy with, the high-frequency, rapidly varying
part of the spectrum; a fact that often results in fractal statistical
processes. For these latter processes the traditional view of deterministic,
predictable signal given by the smooth part of the time series, on which
random, unpredictable noise is superimposed, distorts the dynamics of the
underlying process. Although the research work of this paper has been done
having in mind the former perspective, namely that the complexity of the system is
of internal origin, and that the slow component $S(t)$ is an external
forcing, that does not affect the internal complexity of the system under
study, its result can be used to asses where the latter perspective applies.
For instance, after separating $S(t)$ from $\xi(t)$ in the most rigorous
and objective way possible, we might observe many complex systems, for
instance the teen birth process of many different counties, and to assess if
it is true or not that those characterized by a more intense $S(t)$ might
lead to a larger deviation of $\zeta(t)$ from ordinary scaling.

The outline of the paper is as follows. In Section II we provide a concise
review of diffusion-based methods used to asses the scaling properties of a
time series, the Diffusion Entropy (DE), the Second Moment (SM), the
Multi-Scaling (MS) analysis and the Direct Assessment of Scaling (DAS), and
a brief review of the Wavelet analysis (WA) and its use to reveal the
features of the signal at different time scales. In Section III, with the
help of computer-generated time series, we discuss how the different
diffusion-based methods perceive the addition of a ``slow component" to the
fluctuations, how to detrend this component (we introduce two different
methods: the wavelet smoothing and the step smoothing) and how the
detrending procedures affect the capability of recovering the scaling
properties of the fluctuations. Section IV addresses the same questions of
Section III, but in the case when a ``periodic component" is added to the
fluctuations. In Section V we draw some conclusions.

\section{Illustration of the analysis tools used in this paper}\label{tool}

To make this paper as self-contained as possible, we devote this section to
a concise review of the methods of analysis that we use. These methods are
the diffusion-based methods, the Diffusion Entropy (DE), the Second Moment
(SM), the Multi-Scaling (MS) analysis and the Direct Assessment of Scaling
(DAS), discussed in Section \ref{diffusionmethods}, and the decomposition of
the time series done with the Wavelet Analysis (WA), as illustrated in
Section \ref{waveletmethod}.

\subsection{The diffusion based methods}\label{tool-diffusion}  

\label{diffusionmethods}

As mentioned in the Introduction, the theoretical backdrop to our discussion
is that we have a stochastic time series denoted by the function $\xi \left(
t\right) .$ This time series is used to generate a diffusion process as
follows 
\begin{equation}
x(t)=\int_{0}^{t}\xi (t^{^{\prime }})dt^{^{\prime }}.  \label{difftrajectory}
\end{equation}
Here $x(t)$ is the position occupied by a random walker at time $t$ and the
function $x(t)$ is a diffusion trajectory. Furthermore, it is evident that
to create a probability density function (pdf) we need many diffusion
trajectories of the type indicated by Eq. (\ref{difftrajectory}), namely, an
ensemble of statistically equivalent trajectories.

The continuum form for the diffusion trajectory leaves us with two main
problems. The first problem is typically that we do not have a continuous
time data, in which case the integral in Eq. (\ref{difftrajectory}) is replaced
by a sum and the time $t$ is discrete. The scaling property of Eq. (\ref
{scalingdefinition}) is usually discussed within a continuous-time
perspective, but since the data are discrete, some caution must be exercised
in moving from one representation to the other. The second, and even more
important problem, is how to create many statistically equivalent
trajectories. This is a delicate issue in time series analysis, where, in
general, it is not possible to have many identical replicas of the system,
with many statistically equivalent series of fluctuations. Typically one has
only a single time series. To solve this problem we use an overlapping
windows technique. If $t$ is a discrete time in the interval $\left[
1,N\right] $, we define $N-t+1$ different diffusion trajectories (number of
members in the ensemble) in the following way

\begin{equation}
x_{k}(t)=\sum_{j=k}^{k+t}\xi _{j},\qquad k=1,2,\dots ,N-t+1.
\label{difftrajectory2}
\end{equation}
This corresponds to initiating a ``window'' (interval) of length $t$ at the
data point $k$, and to aggregating all the data in the sequence $\xi _{j}$.
For any window position we sum all the values of the variable $\xi _{j}$
spanned by the window. This procedure is shared by all the diffusion based
methods.

Let us now explore, briefly, the different methods based on creating a
diffusion process from data via Eq. (\ref{difftrajectory2}).

\subsubsection{Diffusion Entropy (DE) Analysis}

The rationale behind the adoption of DE analysis is that, if the pdf of the
diffusion process satisfies Eq. (\ref{scalingdefinition}), then, regardless 
of the pdf shape, the Shannon entropy satisfies the following
relationship

\begin{equation}
S(t)=-\int_{-\infty }^{\infty }p(x,t) \ln\left[ p(x,t)\right] dx=A+\delta 
\text{ } \ln(t),  \label{derelation}
\end{equation}
where $A$ is a constant defined by 
\begin{equation}
A=-\int_{-\infty }^{\infty }F(y) \ln\left[ F(y)\right] dy,\qquad y=\frac{x}{%
t^{\delta }}.  \label{Adefinition}
\end{equation}
Therefore the scaling condition of Eq. (\ref{scalingdefinition}) can be detected
by searching for the linear dependence of the entropy of Eq.(\ref
{derelation}) on a logarithmic time scale.

In practice, the numerical procedure necessary to evaluate the pdf
requires that the x-axis be divided into cells of a given size,
that, in principle, might also depend on the cell position on the x-axis. We
adopt the criterion of assigning to each cell the same size, but we leave
this size time-dependent. For this reason we denote the cell size with the symbol $%
\Delta \left(t\right)$. At any time $t$, the size $\Delta \left(t\right)$
must be chosen so as to lead to a fair approximation of $p\left( x,t\right)$, through the histogram relation

\begin{equation}
p\left( x_{j},t\right) \approx \frac{P_{j}}{\Delta \left( t\right) }.
\label{new1}
\end{equation}
Here $p\left( x_{j},t\right) $ is the histogram at the value $x_{j}$, the
midpoint of the $j$-th cell, at time $t$ and $P_{j}$ is the fraction of the 
total number of trajectories found in this cell at time $t$. The
rationale behind the choice of a time-dependent size for the cell has to do
with obtaining a good estimate of the pdf from the histogram. At early
times, when the trajectories are close together, a constant $\Delta $ is
adequate to estimate the ratio $P_{j}/\Delta .$ As the trajectories diffuse
apart, however, either more trajectories are needed, or a larger $\Delta $
is needed, to provide a reasonable estimate for this ratio. To ensure that (%
\ref{new1}) is satisfied at any time we evaluate the standard deviation of
the diffusion process, $\sigma \left( t\right) $, and select the
cell size to be a fraction of the standard deviation, $\Delta \left(
t\right) =\epsilon \sigma \left( t\right) $ where $0.1\leq \epsilon $ $\leq
0.2.$ There is some sensitivity to the choice of $\epsilon$, but in the
proper range of values in which Eq. (\ref{new1}) is satisfied, the diffusion
entropy

\begin{equation}
S\left( t\right) =-\int_{-\infty }^{\infty }p(x,t)ln\left[ p(x,t)\right]
dx\approx -\sum_{j}P_{j}\ln P_{j}+\ln \Delta \left( t\right)   \label{new2}
\end{equation}
is insensitive to the particular fraction of the standard deviation
adopted.

It has to be pointed out that the DE analysis can yield a logarithmic
dependence on time even when the scaling condition of Eq. (\ref
{scalingdefinition}) is not fulfilled. This is the case of the symmetric L%
\'{e}vy walk studied by the authors of \cite{de7}. L\'{e}vy walks \cite
{fundamental} are characterized by the fact that the time for the walker to
cover a given distance is proportional to the distance itself. This
restriction generates a diffusion process limited by diffusion fronts moving
ballistically. In \cite{de7} a case where the population of the ballistic
fronts $P_{peaks}(t)$ decrease extremely slowly, $P_{peaks}(t)\propto
t^{-\beta }$ with $0<\beta <1$, is examined and in this case the pdf between
the ballistic peaks tends to become a L\'{e}vy distribution with scaling
parameter $\delta =\frac{1}{\beta +1}$. Due to the slow decrease to zero of $%
P_{peaks}(t)$, the diffusion process is bi-scaling (the two scaling
parameters being that of L\'{e}vy for the central part and
the ballistic one for the fronts). The numerical limitation in
creating a perfect power law, makes the DE growing, after a transient,
linearly with a slope equal to $\frac{1}{\beta +1}$ giving the impression
that the condition Eq. (\ref{scalingdefinition}) is fulfilled. Therefore, it is
convenient to denote the slope of the linear dependence on the logarithm of
time of the DE, with the symbol $\delta _{de}$.

\subsubsection{Second Moment (SM) Analysis}

The SM analysis is based on evaluating the standard deviation of the
diffusion process 
\begin{equation}
\sigma (t)=\left[ <\left\{ x(t)-\bar{x}(t)\right\} ^{2}>\right] ^{\frac{1}{2}%
},  \label{smdefinition}
\end{equation}
where $\bar{x}(t)\equiv <x(t)>$ and $<\dots >$ denotes the average over the
ensemble of realizations of diffusion trajectories. If the time
series fluctuations have zero mean value, the quantity $\bar{x}(t)$ vanishes
and Eq. (\ref{smdefinition}) can be written as 
\begin{equation}
\sigma (t)=\left[ \int_{-\infty }^{\infty }x^{2}p(x,t)dx\right] ^{\frac{1}{2}%
},  \label{smrelation}
\end{equation}
which is, in fact, the square-root of the second moment of the distribution.

We evaluate the standard deviation from Eq. (\ref{smdefinition}), and we look
for a time domain where 
\begin{equation}
\sigma (t)\propto t^{\delta _{sm}}\Leftrightarrow \ln\left[ \sigma (t)\right]
\propto \delta _{sm}\text{ }\ln(t).  \label{smscaling}
\end{equation}
is satisfied.  Then, in the time domain for which Eq.(\ref{smdefinition}) is
fulfilled, the standard deviation rescales with the scaling parameter $%
\delta _{sm}$. This scaling does not automatically imply, as
mentioned in Section I, that the proper scaling condition of Eq. (\ref
{scalingdefinition}) is fulfilled.

\subsubsection{Multiscaling (MS) Analysis}

The MS analysis is a generalization of the SM analysis. In fact Eq. (\ref
{smdefinition}) is replaced by the expression 
\begin{equation}
\sigma _{q}(t)=\left[ <|x(t)-\bar{x}(t)|^{q}>\right] ^{\frac{1}{q}},
\label{mscdefinition}
\end{equation}
where $q$ is a real number. It is evident that Eq. (\ref{smdefinition}) is
recovered from Eq. (\ref{mscdefinition}) by setting $q=2$. In case of fluctuations with zero mean value, Eq. (\ref{mscdefinition}) reads:
\begin{equation}
\sigma _{q}(t)=\left[ \int_{-\infty }^{\infty }|x|^{q}p(x,t)dx\right] ^{%
\frac{1}{q}}.  \label{mscrelation}
\end{equation}
If the scaling condition of Eq. (\ref{scalingdefinition}) holds, we can express (%
\ref{mscrelation}) as 
\begin{equation}
\sigma _{q}(t)=B_{q}\text{ }t^{\delta },  \label{mscscaling}
\end{equation}
where 
\begin{equation}
B_{q}=\left[ \int_{-\infty }^{\infty }|y|^{q}F(y)dy\right] ^{\frac{1}{q}}.
\label{deifnitionofbq}
\end{equation}
Note that even in the case where $F(y)$ has a long tail with an inverse
power law index $\mu $, with $\mu <3$, and consequently a divergent second
moment, the relation of Eq. (\ref{mscscaling}) holds true, if $q<2$.

In the case of fractional Brownian motion \cite{mandelbrot}, the
function $F(y)$ is Gaussian, the scaling exponent, usually called $H$ in
this case, in honor of Hurst, is a number in the interval $[0,1]$
and 
\begin{equation}
\sigma _{q}(t)\propto t^{H}\qquad \forall q\in ]-1,+\infty ).
\label{mscscalingFBM}
\end{equation}
In the case of a L\'{e}vy flight \cite{kolmo} the function $F(y)$ is a
L\'{e}vy stable distribution of index $\alpha $ \cite{fundamental}. Note
that the index $\alpha $ is related to the inverse power-law index of the
distribution by the relation $\alpha =\mu -1$, and that the scaling $\delta $
is given in this case by $\delta =1/\alpha =1/(\mu -1)$. Thus, in this case,
it is straightforward to prove 
\begin{equation}
\sigma _{q}(t)\propto t^{\frac{1}{\alpha }}\qquad \forall q\in ]-1,\alpha [.
\label{mscscalingLevy}
\end{equation}

As for the SM analysis we look for a time domain where the f$q$-th fractional standard deviation $\sigma_{q}(t)$ rescales with exponent $\zeta(q)$, namely
\begin{equation}
\left[\sigma_{q} (t)\right]^{q} \propto t^{\zeta (q)}\Leftrightarrow \ln \left[(\sigma_{q}(t))^{q}\right] \propto
\zeta \left( q\right) \ln(t).  \label{mscscaling2}
\end{equation}
When the scaling condition of Eq. (\ref{scalingdefinition}) applies: $q$ $\zeta \left( q\right) =\delta q$
. In the ideal case where divergent moments are involved, as we have seen earlier, we should
 limit our observation to values of $q$ smaller than a
given $q_{max}$. In practice, we do the calculation for the entire range of
values from $q=-1$ to $q=\infty $ (in reality the behavior of $\sigma_{q}(t)$ for very high value of $q$, $q>10$ for example, cannot be trusted because they are dominated by rare events and therefore subject to the problem of lack of statistics). This is done because, in real data there
are no moment divergences, due to the fact that time series are of finite
length. We, then, plot the calculated $\zeta (q)$ as a function
of $q$ and see if a straight line results. The corresponding slope is the genuine value of the scaling parameter $%
\delta $, assuming that the scaling condition Eq. (\ref{scalingdefinition})
applies. Of course, with this technique, as in the case of the DE analysis,
there is no compelling reason to ensure that the resulting scaling parameter
corresponds to a genuine scaling property. If the scaling condition exists,
the DE analysis and the MS analysis detect the real scaling parameter (a
property that the SM analysis does not share). The reverse is not
necessarily true.

\subsubsection{Direct Assessment Scaling (DAS)}

The definition of scaling given by Eq. (\ref{scalingdefinition}) has an
attractive physical meaning. Given any two times $t_{2}$ and $t_{1}$, with $%
t_{2}>t_{1}$, the pdf at time $t_{2}$ coincides with the pdf at time $t_{1}$
if the the following self-affine transformation is applied: take the scale
of the variable $x$, and ``squeeze" it by a factor $R=[t_{1}/t_{2}]^{%
\delta }$, simultaneously take the scale of the distribution intensity, and
``enhance" it by the factor $1/R$. In other words, when Eq. (\ref
{scalingdefinition}) is satisfied, the diffusion process is invariant under
this self-affine transformation. This property is the expression of a kind
of thermodynamic equilibrium reached by the complex system under study; one
that is different from that usually obtained if $\delta \neq 0.5$ and $F(y)$
is not Gaussian. The DAS analysis consists exactly of this procedure of
``squeezing" and ``enhancing" aimed at establishing the invariance of the
pdf by a scaling transformation.

The reader might wonder why no significant use is made of the DAS in
literature. The reason seems to be that the detection of real scaling would
necessitate many trials before discovering the right scaling index, if it
exists. We think that the adoption of the DAS only becomes useful after the
DE and MS methods are applied. In fact, those analysis define a possible
candidate for the true scaling. However, these techniques, as we have seen,
do not provide any compelling reason to prove that the resulting scaling is
a genuine property of the system dynamics. Herein, we apply the DE
and the MS analysis and use the DAS to check if we have determined a genuine
scaling or not.

\subsection{Wavelet Analysis (WA)}\label{tool-wavelet}

\label{waveletmethod}

The wavelet transformation \cite{wavelet1,wavelet2} is close in spirit to
the Fourier transformation, but has a significant advantage. The Fourier
transformation decomposes a time series into a superposition of oscillating
modes, each of which lasts forever. The wavelet transformation decomposes
the time series into ``notes" or wavelets \cite{wavelet2}, localized in
time and in frequency. Formally, the wavelet transform $\tilde{f}(s,t)$ of
the function $f(u)$ is defined as follows 
\begin{equation}
\tilde{f}(s,t)\equiv \int_{-\infty }^{+\infty }du:|s|^{-p} \psi^{*}(\frac{%
u-t}{s})f(u),  \label{wavelettr}
\end{equation}
where the symbol $\psi^{*}$ denotes the complex conjugate of the wavelet $%
\psi (u)$. The wavelet $\psi (u)$ is a filter function satisfying the
particular condition, known as the ``admissibility condition" \cite
{wavelet1,wavelet2}. The parameters $s$, $t$ and $p$, are real numbers.
Eq. (\ref{wavelettr}) shows the advantages of the wavelet transformation
in resolving local features of the time series. In fact, the wavelet
transformation rests on a convolution of the signal with the wavelet
rescaled, through the use of the parameter $s$, and centered on $u=t$.
Therefore the parameter $s$ localizes the frequency domain and the parameter 
$t$ localizes the time domain.

Herein we use the discrete version of the wavelet transformation, known as
the Maximum Overlap Discrete Wavelet Transform (MODWT) that gives birth to a
decomposition of the signal in terms of ``approximations" and ``details"
relative to different time scales. Consider the integers $N$ and $K$. The
former is the length of the series under study and $K$ is the greatest
number satisfying the condition $2^{K}<N$. In this condition we can consider 
$K$ different time scales, given by $2^{1},2^{2},\dots ,2^{K}$. Then,
starting from the smallest time scale, the wavelet transformation is used to
divide the original time series $S$ into two components 
\begin{equation}
S=A_{1}+D_{1}.  \label{firststepdec}
\end{equation}
The component $A_{1}$ contains features having a characteristic time scale
greater than $2^{1}$, since a local average of the time series over all time
scales inferior to $2^{1}$ has been performed. Therefore the component $A_{1}
$, often referred as ``approximation", is ``$2^{1}$-smooth", in the sense
that $A_{1}$ can be considered a slow changing function with respect the
time scale $2^{1}$. The component $D_{1}$, on the other hand, contains all
the features of the time series with a characteristic time scale less than $%
2^{1}$ and therefore is called the ``detail". The partitioning procedure
described above can then applied to the approximation $A_{1}$, splitting it
into the functions $A_{2}$ and $D_{2}$. The latter two functions represent,
respectively, the features of the time series with time scales
greater than $2^{2}$ and those features with a time scale less than $2^{2}$,
but greater than $2^{1}$, that have already been removed. Clearly, at the
k-th step of this procedure, with $k<K$, we have the decomposition 
\begin{equation}
S=A_{k}+D_{k}+D_{k-1}+\dots +D_{1},  \label{kstepdec}
\end{equation}
with $A_{k}$ representing the smooth time series referring to the time scale 
$2^{k}$ and $D_{j}$, with $1<j<k$, the detail of time series with the time
scale located in the interval $[2^{k-1},2^{k}]$.

\section{Computer-generated Data: Effects of a Slow Component}\label{cpu-slowcomponent}

In Section I we mentioned that the main goal of this paper is to detect the 
scaling index as a measure of complexity of time series: this task
is confounded by the presence of slow components. In this section we
illustrate the influence of a slow component on a computer-generated random
time series created for this specific purpose. We apply the diffusion
entropy (DE), the second moment (SM), the multiscaling analysis (MS) and the
direct assessment scaling (DAS) analysis to this computer-generated time
series, and establish that it is not possible to reach a reliable conclusion
about the real complexity of the process with any of these techniques.
Consequently we show how to eliminate the influence of the slow component,
without distorting the scaling properties of the fluctuations, if any
exists. This detrending procedure aims at establishing a stationary
condition, to which DE, SM and MS analysis can be productively applied.

Let us introduce the computer-generated time series we use to test
the above processing techniques. We adopt a discrete-time picture, so that the
time series to analyze reads as follows: 
\begin{equation}
\xi _{j}=S_{j}^{T}+\zeta _{j},  \label{slow+random}
\end{equation}
where both $S_{j}^{T}$ and $\zeta _{j}$ are functions of the discrete time 
$j$. The slow component of the time series, $S_{j}^{T}$, is a function,
either regular or stochastic, that changes over a long time scale denoted by 
$T$. The function $\zeta _{j}$ represents the fluctuations and we refer to
it as noise. In the numerical calculations, we assume the noise to
not have any memory. However, the conclusion reached in this section can be
easily extended to the case where $\zeta _{j}$ has a long-time memory, yielding 
a scaling index  $\delta $ moderately larger than the
simple diffusion scaling, $\delta =0.5$. We shall consider three different
kinds of slow components, all of them with zero mean: a linear drift with a
small slope, $SL_{j}^{T}$, a slowly changing continuous function, $SC_{j}^{T}
$, and a step function, $SS_{j}^{T}$. Fig. \ref{figure1} shows these three different
slow components. We select $\zeta $ to be a Gaussian random process
with zero average and standard deviation $\sigma =12$, to which we will
refer as Gaussian noise (GN). This choice of $\sigma $ makes the
intensity of the GN, measured by the standard deviation, comparable
to the intensity of the slow components $SC_{j}^{T}$ and $SS_{j}^{T}$, while
keeping the noise at a slightly larger intensity than that of the $SL_{j}^{T}
$ component.

We note that DE analysis applied to the computer-generated time series,
corresponding to GN, is a perfectly straight line, with $\delta =0.5$. The
use of random, rather than correlated, fluctuations in the simulation is
dictated by simplicity. This choice of a memoryless noise makes the
non-stationary effects, produced by the slowly changing first term on the
right-hand side of Eq. (\ref{slow+random}), more transparent. However, the
theoretical arguments adopted, in the remaining part of this section, to
explain these significant effects, are independent of whether the noise is
memoryless or correlated. Finally, the length, $13149$, of the data has
been chosen for ease of application of these results in paper II.

\subsection{Diffusion Entropy (DE) and Second Moment (SM) Analysis}

In this section we apply the DE and the SM analysis to the
computer-generated sequences of Eq. (\ref{slow+random}). Figs. \ref{figure2} and \ref{figure3}
illustrate the DE and the SM methods, respectively. In both figures the
top, middle and bottom frame refer to $SS_{j}^{T}$, $SC_{j}^{T}$and $%
SL_{j}^{T}$, respectively. In each frame we compare the results relative to
the application of the methods to the time series defined by the sum of the
slow component and the GN, as in Eq. (\ref{slow+random}), with the results
obtained when the two components are considered separately.

Regarding the slow components $SS_{j}^{T}$and $SC_{j}^{T}$, we see from Fig. \ref{figure2}
that at short times the DE generated by the time series of Eq. (\ref{slow+random}%
), is different from both the DE of the slow component and the DE of the GN
component. This is explained by the fact that the slow component and GN have
comparable intensities. Thus, it is evident that their joint action
generates a faster spreading of the pdf and therefore a faster entropy
increase. At long times the joint action of the two components yields the
same effects as the slow component alone. This is due to the fact that the
slow component generates ballistic diffusion, which is faster than the
simple diffusion generated by the GN alone.

A different behavior appears with the slow component $SL_{j}^{T}$. In this
case, the short-time behavior is dominated by the noise component. This
dominance by the noise occurs because the intensity of this component is
smaller than the noise intensity. However, even in this case the long-time
behavior is dominated, as explained earlier, by the slow component. This is a remarkable
property, because in this case a mere visual inspection of the time series
is not sufficient to reveal the presence of a bias, which is hidden by very
large fluctuations. In any event, the adoption of the DE method yields a
large scaling exponent that one might erroneously attribute to highly
correlated fluctuations. This is an effect that we have to take into account
when analyzing real time series.

The SM analysis reveals properties similar to those emerging from the DE
analysis, the only relevant difference is that the convergence to the steady
condition of the slow component alone is much faster than in the
corresponding case of the DE analysis. It is worthwhile to discuss this
result in detail. Let us define $\sigma _{tot}(t)$ as the total standard
deviation, at a time $t$, of the diffusion process relative to the sum of
the slow component and the noise. Under the assumption that the individual
components of the signal of Eq. (\ref{slow+random}) are independent of one
another, we write

\begin{equation}
\sigma _{tot}^{2}(t)=\sigma _{slow}^{2}(t)+\sigma _{noise}^{2}(t)=\sigma
_{slow}^{2}(t)\left[ 1+\frac{\sigma _{noise}^{2}(t)}{\sigma _{slow}^{2}(t)}%
\right] .  \label{sigmadecomposition}
\end{equation}
The SM analysis rests on evaluating the increase with time of the logarithm
of the standard deviation. With some elementary algebra, we convert Eq. (\ref
{sigmadecomposition}) into 
\begin{equation}
\log \left[ \sigma _{tot}(t)\right] =\log \left[ \sigma _{slow}(t)\right]
+ \frac{1}{2} \, \log \left[ 1+\frac{\sigma _{noise}^{2}(t)}{\sigma _{slow}^{2}(t)}\right] .
\label{logsigmadecomposition}
\end{equation}
When the noise standard deviation is smaller than the slow component
standard deviation, namely, $\sigma _{noise}(t)\ll \sigma _{slow}(t)$, using
the Taylor series expansion of the logarithm. we obtain 
\begin{equation}
\log \left[ \sigma _{tot}(t)\right] \approx \log \left[ \sigma
_{slow}(t)\right] + \frac{1}{2} \, \frac{\sigma _{noise}^{2}(t)}{\sigma _{slow}^{2}(t)}.
\label{logsigmaapproximation}
\end{equation}
Therefore, when $\sigma _{noise}(t)\ll \sigma _{slow}(t)$ the leading
contribution of the SM analysis is the logarithm of the slow component
standard deviation, and the next expansion term is the square of the ratio
of the noise standard deviation to the slow component standard deviation. In
the case of diffusion entropy the numerical results indicate that a
plausible expression to use is

\begin{equation}
S_{tot}(t)=S_{slow}(t)+O\left[ \frac{\sigma _{noise}(t)}{\sigma _{slow}(t)}%
\right] ^{\alpha },  \label{entropyapproximation}
\end{equation}
with $\alpha <2$. In fact, the numerical results illustrated in the frames
of Fig. \ref{figure2} corresponding to $SC_{j}^{T}$and $SS_{j}^{T}$reveal that the
diffusion entropy is more sensitive to the noise component than is the SM
analysis. This sensitivity is due to the convergence to the behavior
dominated by the slow component being faster than the one relative to the SM analysis. For
a deeper understanding of the diffusion created by the superposition of the noise and slow component, in the next subsection we use
the multiscaling analysis.

\subsection{Multiscale (MS) Analysis}\label{ms-application}

In Fig. \ref{figure4} we observe that in the long-time limit, $t>100$, the DE produced
by the time series resulting from the sum of noise and slow component $SS_{j}^{T}$ coincides with the DE generated by the slow component alone that
increases with a slope equal to $1$. This asymptotic property has already
been discussed. Our goal here is to shed light on the convergence to the
long-time scaling of the whole time series, with the help of multiscaling
analysis. To accomplish this goal, we divide the time range, explored by the
DE analysis, into three time regions: times smaller than $10$ (early
stage of the diffusion process), times between $10$ and $100$ (middle
stage of the diffusion process) and times from $100$ to $1000$ (later
stage of the diffusion process). We apply the multiscaling analysis to the
two time series in Fig. \ref{figure4}, for each of these three time regions, as shown in
Fig. \ref{figure5}.

The three frames of Fig. \ref{figure5}, from bottom to top refer to the early, the middle
and the later stage of the diffusion process, respectively. The squares
denote the numerical results relative to the slow component alone, the
triangles denote the numerical results concerning the sum of slow component
and noise and the dashed line corresponds to a straight line of slope $1$,
as expected for a diffusion rescaling ballistically. Moving from the bottom
to the top we notice that the agreement between triangles and squares tend
to increase. At the top, there is an almost complete equivalence throughout
the whole $q$-region explored. We also see that in the early and middle
region the disagreement between triangles and squares tend to increase upon
increasing $q$. This means that the difference between the two cases becomes
more and more significant at larger and larger distances. The presence of
noise tends to slow down the distribution broadening.

To understand the influence of the GN on the slow component, we notice that
the position of any of the diffusing trajectories can be written at
time $t$ as 
\begin{equation}
x(t)=x_{slow}(t)+x_{noise}(t),  \label{postrajectory}
\end{equation}
where the contributions to $x(t)$ are separated into the slow component and
noise, respectively. Now, since the noise has a vanishing mean, we have $50\%
$ probability that the absolute value of $x(t)$ is increased by the presence
of the noise and $50\%$ that it is decreased. When we raise the absolute
value of $x(t)$ to a power of $q$ larger than $1$, the half with positive
increment contribute to the q-th moment with a greater weight than the other
half. So, at any given time, the presence of the noise component makes the
q-th moments , with $q>1$, larger than in the case without noise, namely,
the larger $q$, the larger the discrepancy. We know that at long times, the
moments of the two distributions, those with and without noise, must
coincide. As earlier pointed out, this occurs, because the noise component
has slower diffusion. Consequently, the moments of the distribution with
noise undergo a slower increase than the moments of the distribution without
noise.

\subsection{Direct Assessment Scaling (DAS)}\label{dasatwork}

Finally, we want to use the DAS method to shed light on the apparently
confusing situation emerging from the use of DE, SM, and MS analysis. We
have seen that the DE suggests that $\delta =0.86$ might be a reasonable
measure of scaling, thereby suggesting the existence of pronounced
cooperation in the system under study. The MS method, on the contrary,
suggests that the ballistic scaling, $\delta =1,$ is a more appropriate
representation of the system complexity, at least in the long-time regime.
Using the DAS we discover that neither of these two conditions is a proper
representation of the system dynamics. Previously, we studied three distinct
time regimes, short, intermediate and long. Here we focus on the
intermediate time regime, ranging from $10$ to $100$. This choice is
dictated by the fact that the deviation from a straight line in the middle
frame of Fig. \ref{figure4} suggests that the DE prediction of $\delta =0.86$ is
questionable.

We apply the DAS method, namely, the squeezing and enhancing technique,
assuming for the scaling parameter $\delta $, the values $1$, $0.86$ and $0.6
$. The corresponding results are illustrated in the top, middle and bottom
frames of Fig. \ref{figure6}, respectively. We see that the adoption of $\delta =1$
leaves the tail and the peak positions unchanged. However, the peak
intensity (in particular the intensity of the central peak) is not
invariant. This lack of invariance of the peaks means that $\delta =1$ is an
acceptable scaling parameter for the skeleton of the pdf and its tail, but
not for the peak$^{\prime }$s intensities. The adoption of $\delta =0.86$,
as suggested by the DE analysis, is satisfactory for both peak intensities
and position. However, this scaling property is limited to the central part
of the histogram. We see from the middle frame of Fig. \ref{figure6} that the choice of
the scaling parameter indicated by DE has difficulty with the side portions
of the histogram. In fact, we see that the peak at $x=-200$ is annihilated
by the adoption of the DAS method. Finally, as we see in the last panel, the
scaling parameter $\delta =0.6$ turns out to be unsatisfactory everywhere.

In summary, the scaling parameter $\delta =0.86$ emerging from DE analysis
depicted in Fig. \ref{figure4} does not reflect the true scaling of the
computer-generated process. Fig \ref{figure6} indicates that the scaling $\delta =0.86,$
afforded by DE, refers to the central part of the distribution, whereas the
sides of the histogram are more satisfactorily represented by the scaling $%
\delta =1$. This is reminiscent of the multiscaling properties revealed,
using the MS analysis , in the paper of Ref. \cite{de7}. However, in the
case studied in Ref. \cite{de7} the scaling of the central part,
corresponding to L\'{e}vy walk dynamics, turned out to be a fair reflection
of the correlated nature of fluctuations. Here, this anomalous scaling is
not the effect of correlated fluctuations, rather it is the consequence of
the non-stationary effects produced by the slow component superposed on the
GN with our computer-generated data.

\subsection{Detrending the Slow Component}

The above results establish that a slow component can produce misleading
effects that have to be removed in order to detect the scaling properties of
stationary fluctuations. The possible correlations remaining after the slow
component has been removed, indicate the amount of cooperation of the
system. In the following subsections we illustrate two distinct detrending
methods, the ``step smoothing" and ``wavelet smoothing``.

\subsubsection{Step Smoothing}

\label{stsm}

\begin{figure}[tbp]
\begin{center}
\includegraphics[angle=270,width=5.5in] {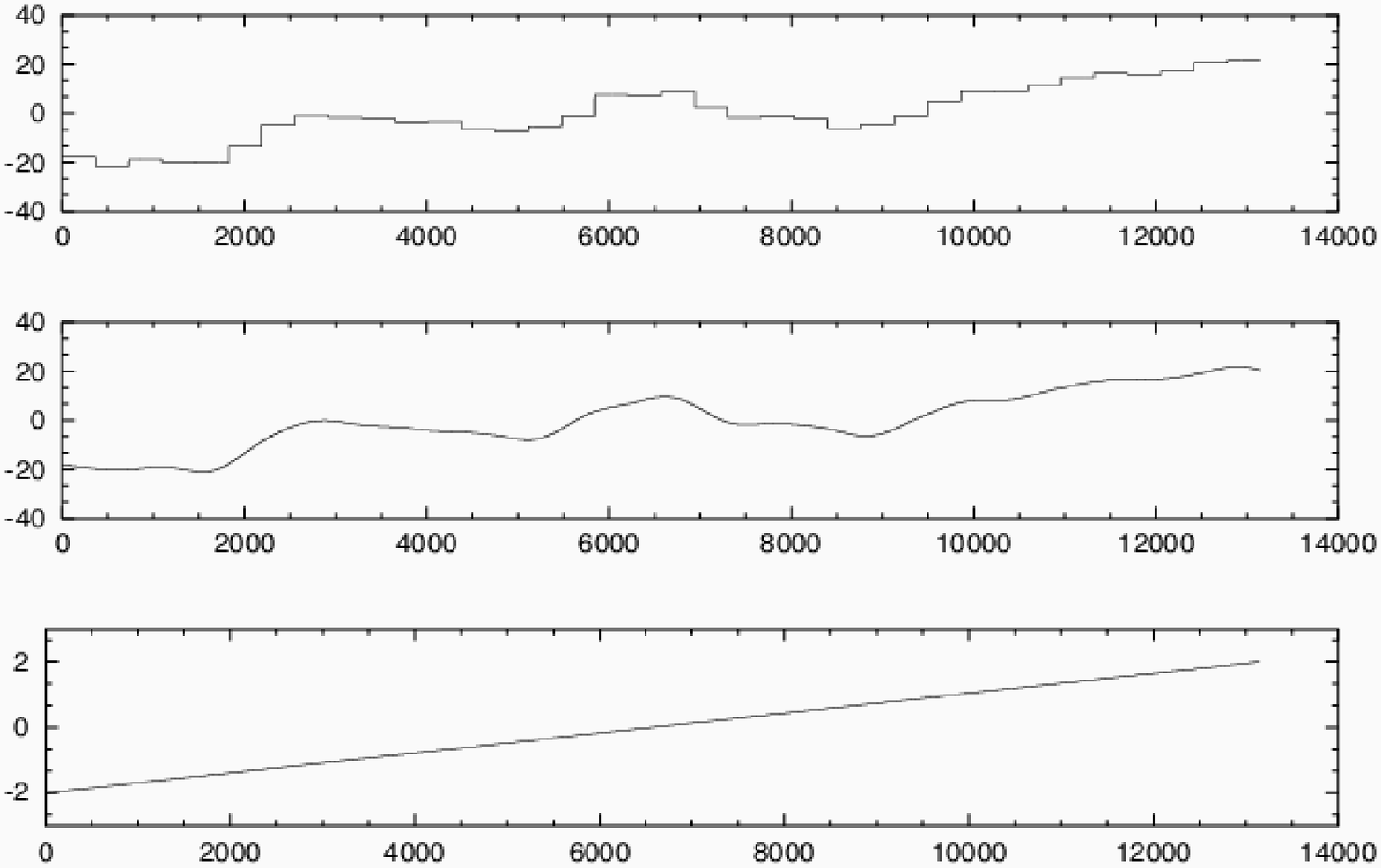}
\caption{The three different slow components adopted. From the top to 
the
bottom frame: step function, continuous smooth function and straight 
line
with a small slope. The initial value of the straight line is $-2$ and 
the
final, at a time $t=13149$, is $2$.}
\label{figure1}
\end{center}
\end{figure}

\begin{figure}[tbp]
\includegraphics[angle=-90,width=5.5in] {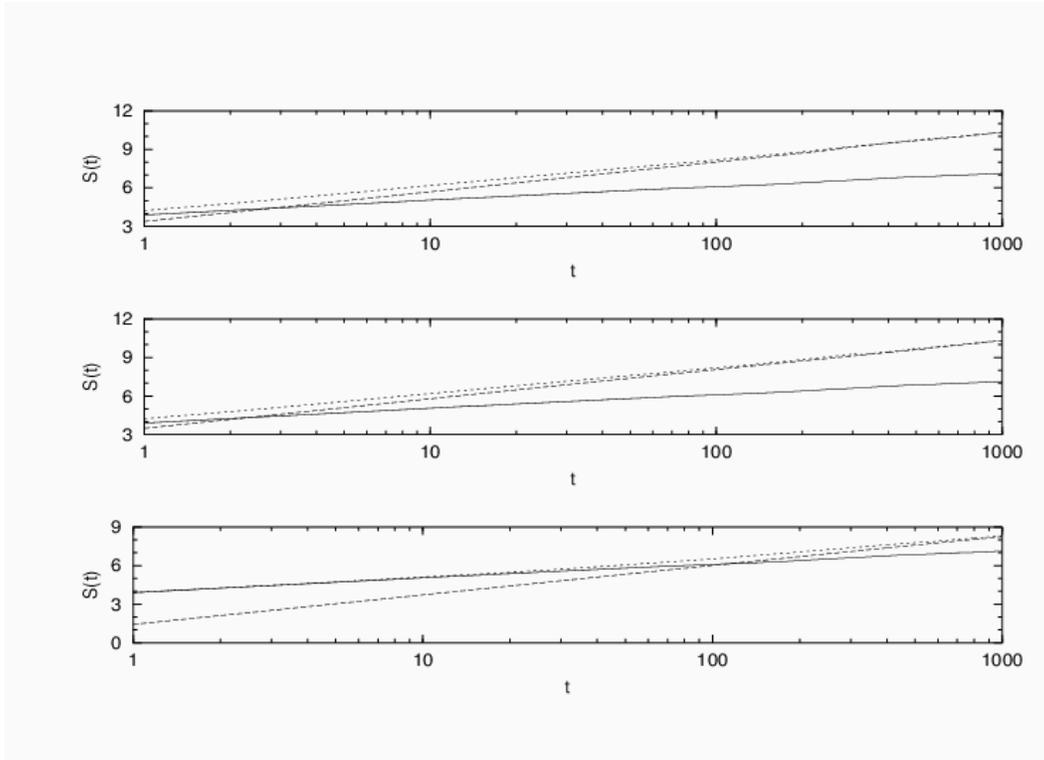}
\caption{The diffusion entropy, $S(t)$, as a function of time $t$, in a
logarithmic scale. Each frame shows three different curves, concerning
noise, solid line, slow component , dashed line, and sum of noise and 
slow
component, dotted line. As far as the slow component is concerned, from 
the
top to the bottom frame we illustrate the results concerning 
$SS_{J}^{T}$, $SC_{J}^{T}$, and $SL_{J}^{T}$, namely the slow components illustrated in 
the
corresponding frames of Fig. 1.}
\label{figure2}
\end{figure}

\begin{figure}[tbp]
\includegraphics[angle=-90,width=5.5in] {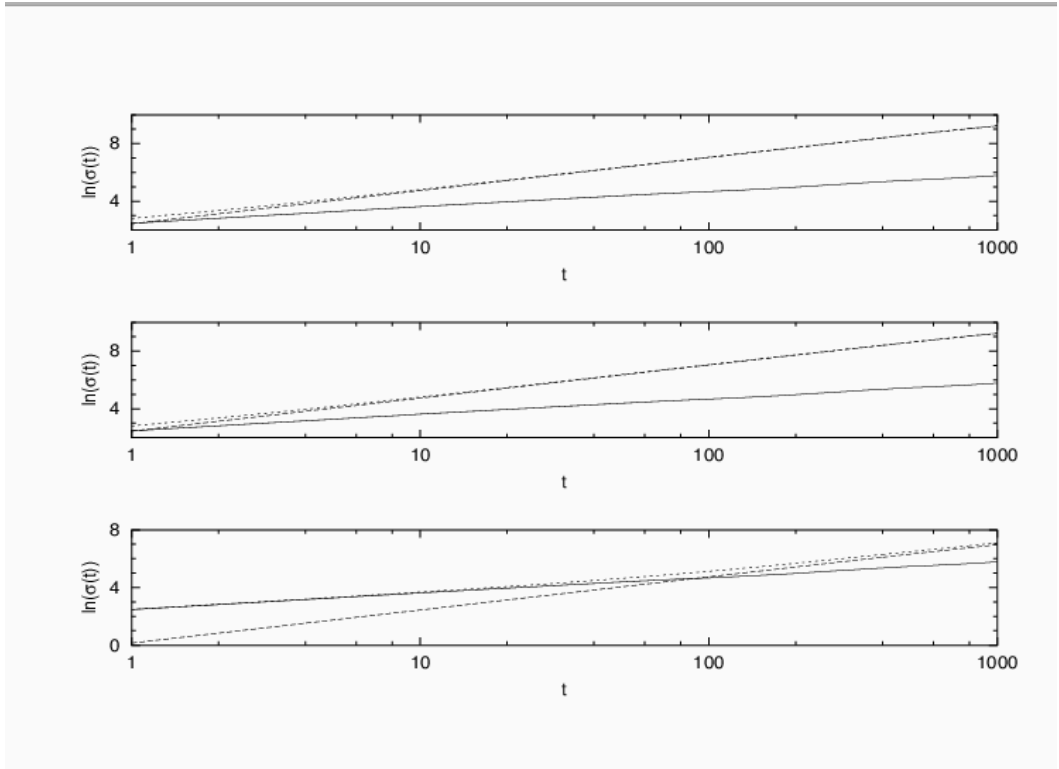}
\caption{The logarithm of the standard deviation, $\ln(\sigma(t))$, as a
function of time $t$, in a logarithmic scale. Each frame contains a set 
of
three curves, the full line referring to the noise alone, the dashed 
line to
the slow component alone and the dotted line to the sum of slow 
component
and noise. From the top to the bottom frame these sets of curves refer 
to $SS_{J}^{T}$, $SC_{J}^{T}$ and $SL_{J}^{T}$, respectively. These are the 
slow
components of the top, middle and bottom frame of Fig. 1, respectively. 
}
\label{figure3}
\end{figure}

\begin{figure}[tbp]
\includegraphics[angle=-90,width=5.5in] {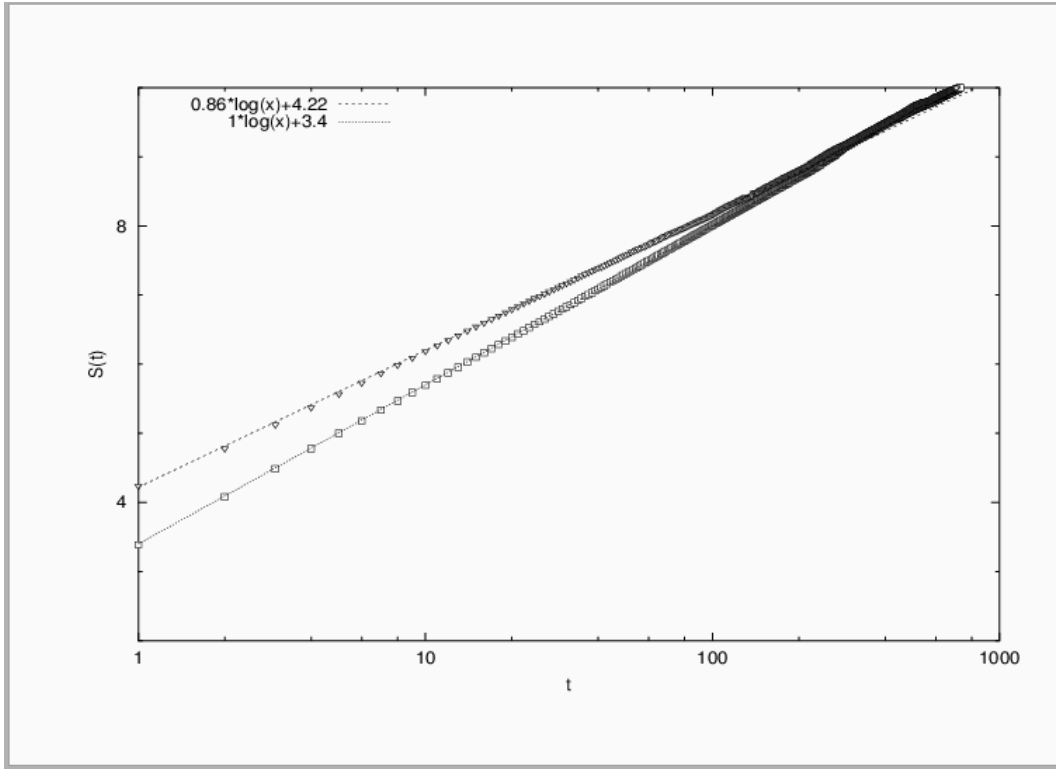}
\caption{The diffusion entropy $S\left( t\right) $ as a function of the
logarithm of time. The squares denote the time series corresponding only 
to the slow component $SS_{j}^{T}$ and the triangles denote the slow 
component $SS_{j}^{T}$ plus noise (bottom frame of Fig. 2). We note that the two
straight lines suggest that in the time range from 10 to 100 the former and
the latter curves correspond to $\delta =0.86$ and $\delta = 1$,
respectively.}
\label{figure4}
\end{figure}

\begin{figure}[tbp]
\includegraphics[angle=-90,width=5.5in] {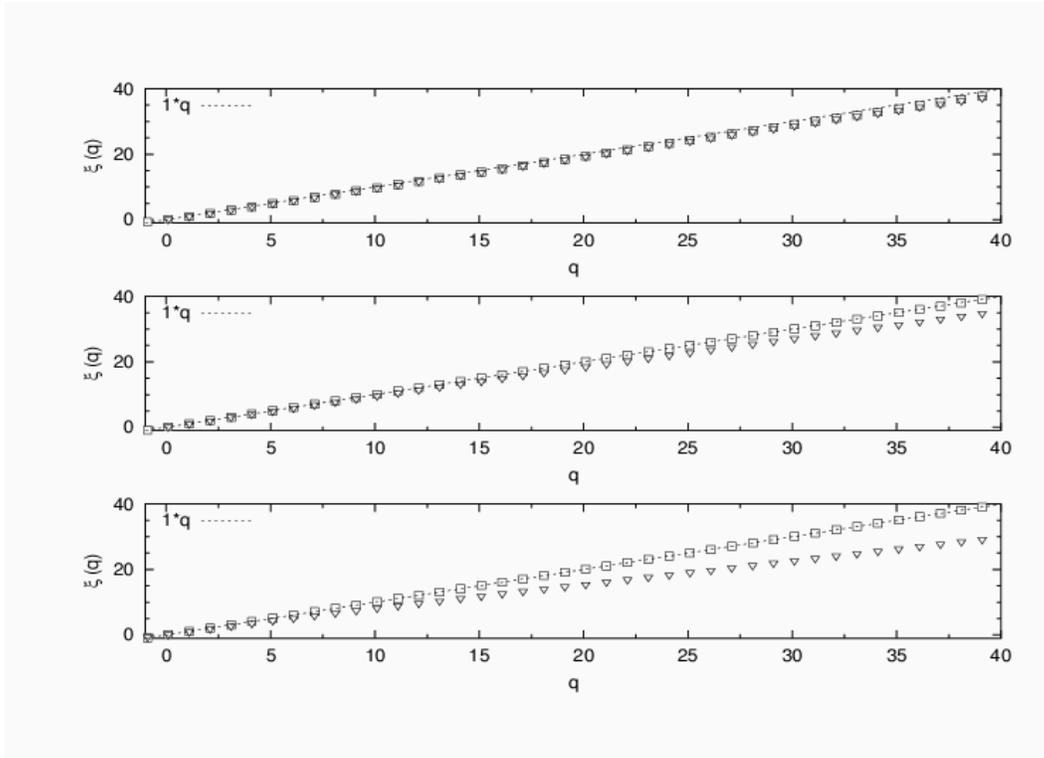}
\caption{The exponent $\xi(q)$ as a function of the parameter $q$. The
squares and the triangles denote the results of the MS analysis applied 
to
the slow component alone and to the the sum of the slow component and 
noise,
respectively. The three frames refer, from the top to the bottom to the
short-time region, betwenn 1 and 10, the middle-time region, betwen 10 
and
100, and the large-time scale, between 100 and and 1000.}
\label{figure5}
\end{figure}

\begin{figure}[tbp]
\includegraphics[angle=-90,width=5.5in] {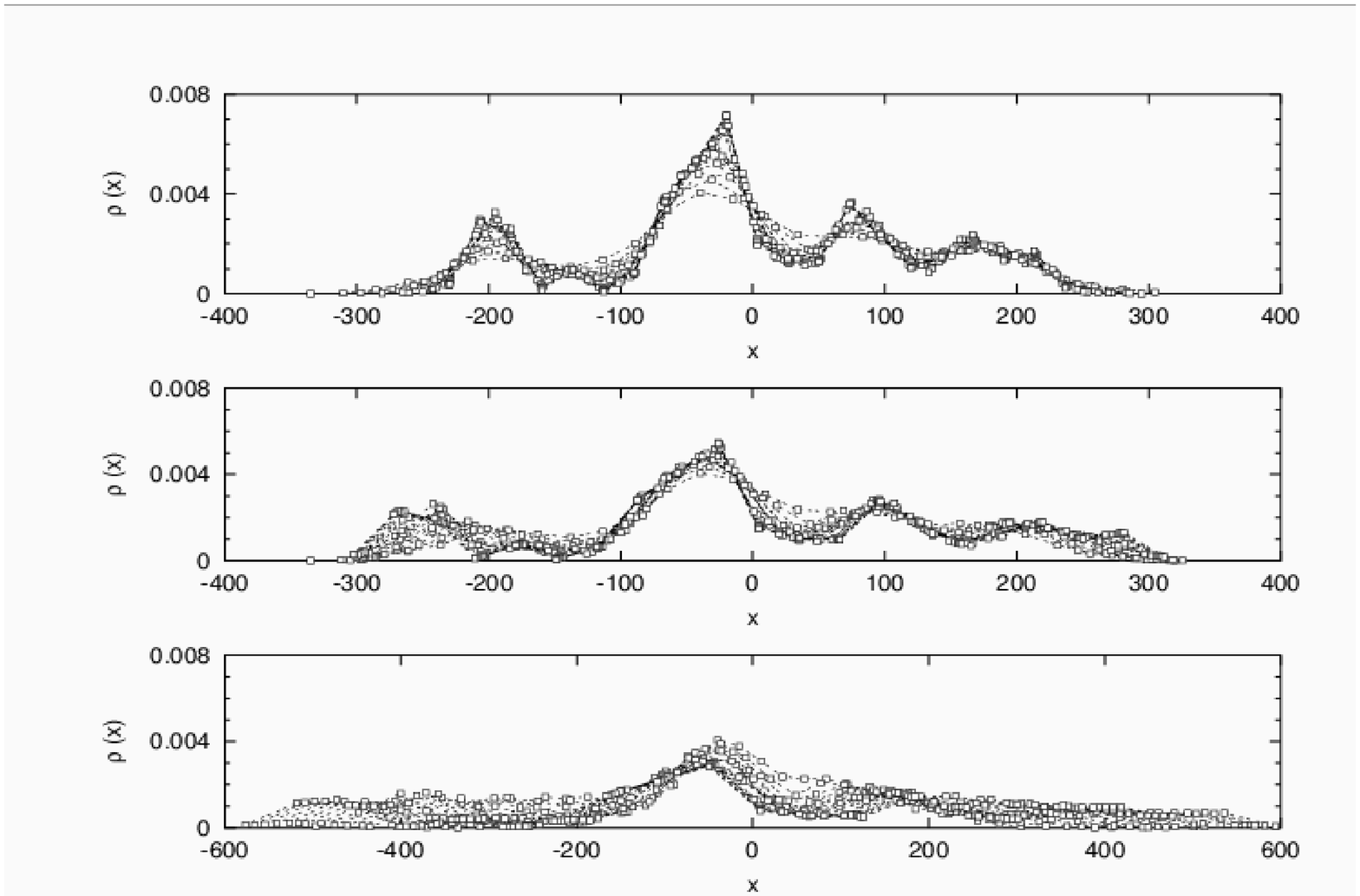}
\caption{The DAS analysis of the time series given by the sum of the slow
components $SS_{j}^{T}$ plus the GN. We consider the middle-time
region, defined in Section \ref{ms-application}. On the axis of the ordinates $\rho
\left( x\right) $ we plot the histograms produced by adopting different
squeezing and enhancing transformations, described in Section \ref{dasatwork}, to
assess to what extent the various histograms coincide. The three frames
refer, from top to bottom, to DAS analysis with the scaling parameter, 
$\delta =1,$ $0.86$ and $0.6,$ respectively.}
\label{figure6}
\end{figure}

The ``step smoothing'' procedure consists of dividing the time series into
non-overlapping patches of length equal to the characteristic time $T$: we
evaluate the average value inside each patch and we subtract it from the
data. In other words, this procedure consists of approximating the slow
component with a step function of the same kind as that of the top frame of
Fig.\ref{figure1}. The details of the procedure are as follows. Start from Eq. (\ref
{slow+random}) and create the new variable $X_{j}$, the sum of the variable $%
\zeta $ inside the $j$-th patch, namely, 
\begin{equation}
X_{j}=\sum_{k=jT}^{k<(j+1)T}\xi _{k}\quad j=0,1,2,..,P,
\label{definitionofxj}
\end{equation}
where $P$ is the number of patches of length $T$ in the time series. To make
explicit the contribution to $X_{j}$ of the two components (the slow and the
random) of the variable $\xi _{j}$ we define

\begin{eqnarray}
R_{j} &=&\sum_{k=jT}^{k<(j+1)T}\zeta _{k}\quad j=0,1,2,...,P
\label{definitionofRjAj} \\
A_{j} &=&\sum_{k=jT}^{k<(j+1)T}S_{k}^{T}\quad j=0,1,2,...,P
\end{eqnarray}
and therefore

\begin{equation}
X_{j}=A_{j}+R_{j}.  \label{twopieceszj}
\end{equation}
If it happens that in any single patch the second term of the right-hand
side of Eq. (\ref{twopieceszj}) is negligible compared to the first term, we can
use this equation to evaluate the average of the slow component and consider
it as a fair approximation to the slow component inside each patch.
Moreover, we notice that the noise component has zero mean. Thus, the most
probable value for $R_{j}$ is zero, and the error associated with this
prediction is given by the standard deviation $\sigma _{R_{j}}$, therefore
 $R_{j}=0\pm \sigma _{R_{j}}$. Then, using the definition Eq. (\ref
{definitionofRjAj}) we apply the following approximation 
\begin{equation}
\sigma _{R_{j}}\approx \sigma _{0}\text{ }T^{\delta _{sm}},
\label{stddevofRj}
\end{equation}
where $\sigma _{0}$ is the standard deviation of the variable $\zeta $ and
the exponent $\delta _{sm}$ is a number between $0$ and $1$. If the scaling
condition applies and the standard deviation is finite, this exponent is the scaling coefficient, that is, $%
\delta _{sm}=\delta $. If the scaling condition does not apply, this
exponent is given by the mean slope. Finally, with the help of Eq. (\ref
{stddevofRj}) we can state that $X_{j}\approx A_{j}$ when 
\begin{equation}
\sigma _{0}\text{ }T^{\delta _{sm}}\ll |a_{j}|\text{ }T\Rightarrow \sigma
_{0}\ll |a_{j}|\text{ }T^{1-\delta _{sm}},  \label{conditionofdetrendability}
\end{equation}
where $a_{j}$ is the average amplitude of the slow component in the j-th
patch, and, with this equation holding true,

\begin{equation}
a_{j}\approx \frac{X_{j}}{T}  \label{practicalBj}
\end{equation}

\begin{figure}[tbp]
\includegraphics[angle=-90,width=5.5in] {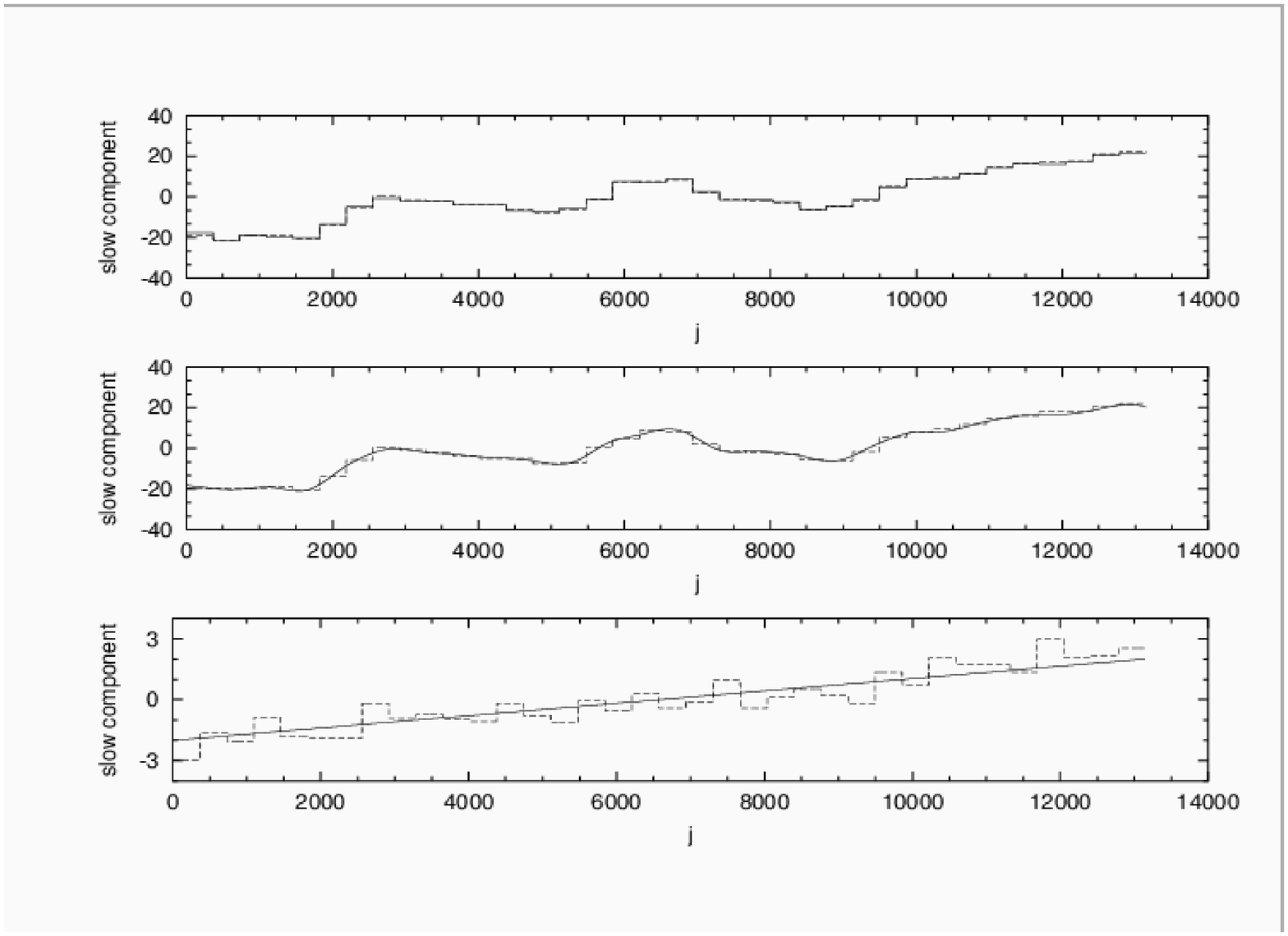}
\caption{The step smoothing technique of Section \ref{stsm} at work. 
With
the full line we indicate the results of the step smothing procedure, 
and
with the dashed line the slow component to derive. From the top to the
bottom frame we refer to the case where the slow component are 
$SS_{j}^{T}$, $SC_{j}^{T}$ and $ST_{j}^{T}$, resspectively .}
\label{figure7}
\end{figure}

In Fig. \ref{figure7} we show the results of this detrending procedure for the three
time series with the three slow components of Fig. \ref{figure1}, with characteristic
time $T=365$ and with GN. We choose the value 365 in anticipation of the
application of these techniques to yearly data in paper II. We see that
the slow components $SS_{j}^{T}$ and $SC_{j}^{T}$ (top and middle frame of
Fig. \ref{figure7}) are fairly well reproduced, while the result for the slow component $%
SL_{j}^{T}$ (bottom frame of Fig. \ref{figure7}) is poor. The reason for this behavior is
that the first two cases satisfy the condition of Eq. (\ref
{conditionofdetrendability}) while the last one does not. In fact, in the
last case the absolute value of the average, $|a_{j}|$, relative to the $j$-th patch
 is considerably smaller than $\sigma _{0}$, the intensity of the GN.

\subsubsection{Wavelet Smoothing}

\label{wavsm}

Another way of obtaining an approximation to the slow component is to use
the chain of approximations and details stemming from the MODWT. If $T$ is
the characteristic time of the slow component, a good approximation is given
by the $j$-th wavelet approximation, the one relative to scale $2^{j}$ where $%
j $ is such that $2^{j}\ $is as close as possible to the time $T$ ($T=365,$
so both $j=8$ and $j=9$ are good choices). In fact, the wavelet
transformation acts as a filter on the contributions corresponding to time
scales smaller than the time scale examined. The detrending procedure rests
on eliminating the component corresponding to the $T$-time scale from the data.

\begin{figure}[tbp]
\includegraphics[angle=-90,width=5.5in] {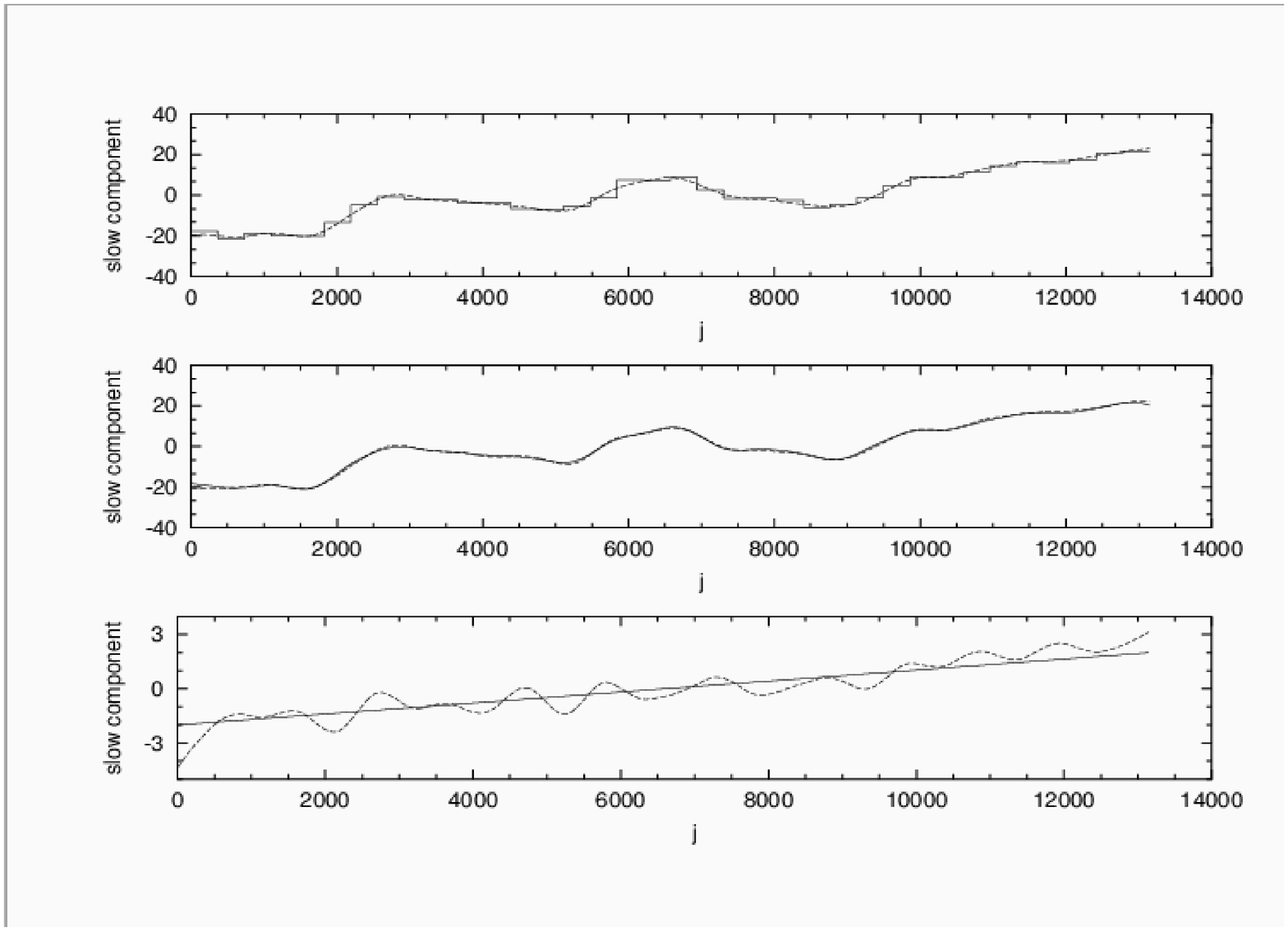}
\caption{ The wavelet smoothing technique of Section \ref{wavsm} wiht 
the
wavelet db8, at work. With the full line we indicate the results of the 
step
smothing procedure, and with the dashed line the slow component to 
derive.
From the top to the bottom frame we refer to the case where the slow
component are $SS_{j}^{T}$, $SC_{j}^{T}$ and $ST_{j}^{T}$, 
respectively.}
\label{figure8}
\end{figure}

\begin{figure}[tbp]
\includegraphics[angle=-90,width=5.5in] {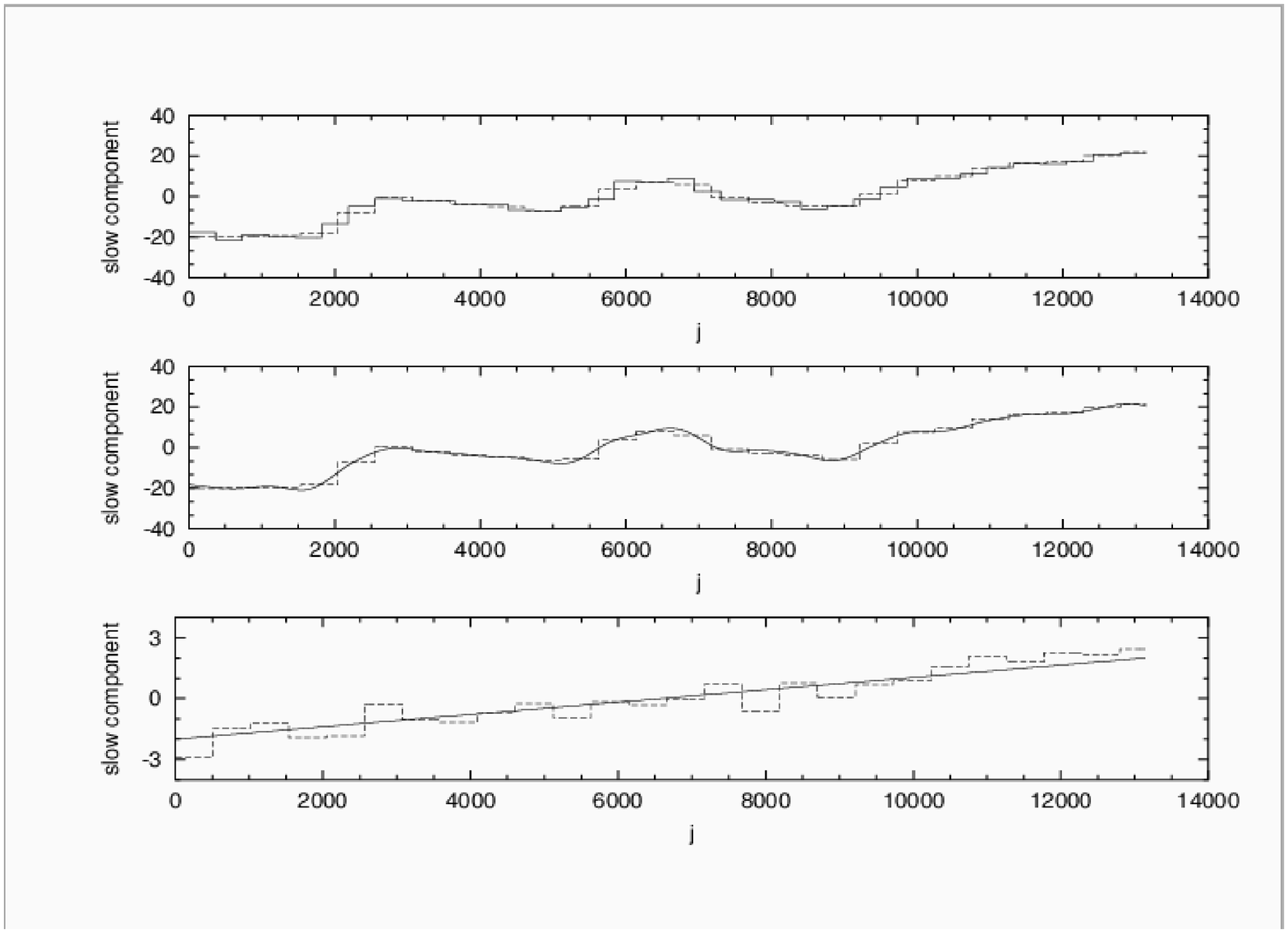}
\caption{The wavelet smoothing technique of Section \ref{wavsm} wiht the
wavelet db1, at work. With the full line we indicate the results of the 
step
smothing procedure, and with the dashed line the slow component to 
derive.
From the top to the bottom frame we refer to the case where the slow
component are $SS_{j}^{T}$, $SC_{j}^{T}$ and $ST_{j}^{T}$, 
respectively.}
\label{figure9}
\end{figure}

We apply the wavelet smoothing to the three time series that have been
analyzed, in Fig. \ref{figure7}, by means of the step detrending procedure. Our purpose
is to show that the two detrending techniques yield virtually the same
results. We choose as analyzing wavelet the Daubechies wavelet number $1$
and $8$, the same as those adopted elsewhere \cite{nicola}. We denote these
wavelets as db1 and db8, respectively. We adopt, as a scale, $2^{9}=512$.
The results are shown in Figs. \ref{figure8} and \ref{figure9}, respectively. In both cases, as
expected, we get results very similar to what obtained with the step
procedure. This means that the wavelet method very satisfactorily reproduces
both the $SS_{j}^{T}$ and $SC_{j}^{T}$ (top and middle frame) time series,
but it turns out to be as inaccurate as the step detrending method, and very
similar as well, when applied to the $SL_{j}^{T}$ time series (bottom frame). 
This equivalence is a consequence of the wavelet result retaining, in
part, the features of the mother wavelet, which is, in turn, a square wave.
The adoption of the db8 wavelet yields a result that is as inaccurate as
that of wavelet db1 and the step detrending method.

\subsubsection{The Effects of the Detrending Procedure}

We are, now, in a position to check a crucial step of our approach to the
search of the complexity of a given process. In the next section we shall
show how to detrend a periodic bias. When this approach is applied to real
sequences in paper II, we derive a time series similar to the slow
component plus noise that we are discussing here. Then, we shall have to
detrend the slow component using the procedure of this subsection. For the
sake of simplicity, the computer-generated time series, here under study,
has been chosen to be the sum of a slow component and uncorrelated
fluctuations. Consequently, in the case here under study the fluctuations
emerging from the adoption of the detrending procedure should be random,
therefore yielding $\delta =0.5$.

\begin{figure}[tbp]
\includegraphics[angle=-90,width=5.5in] {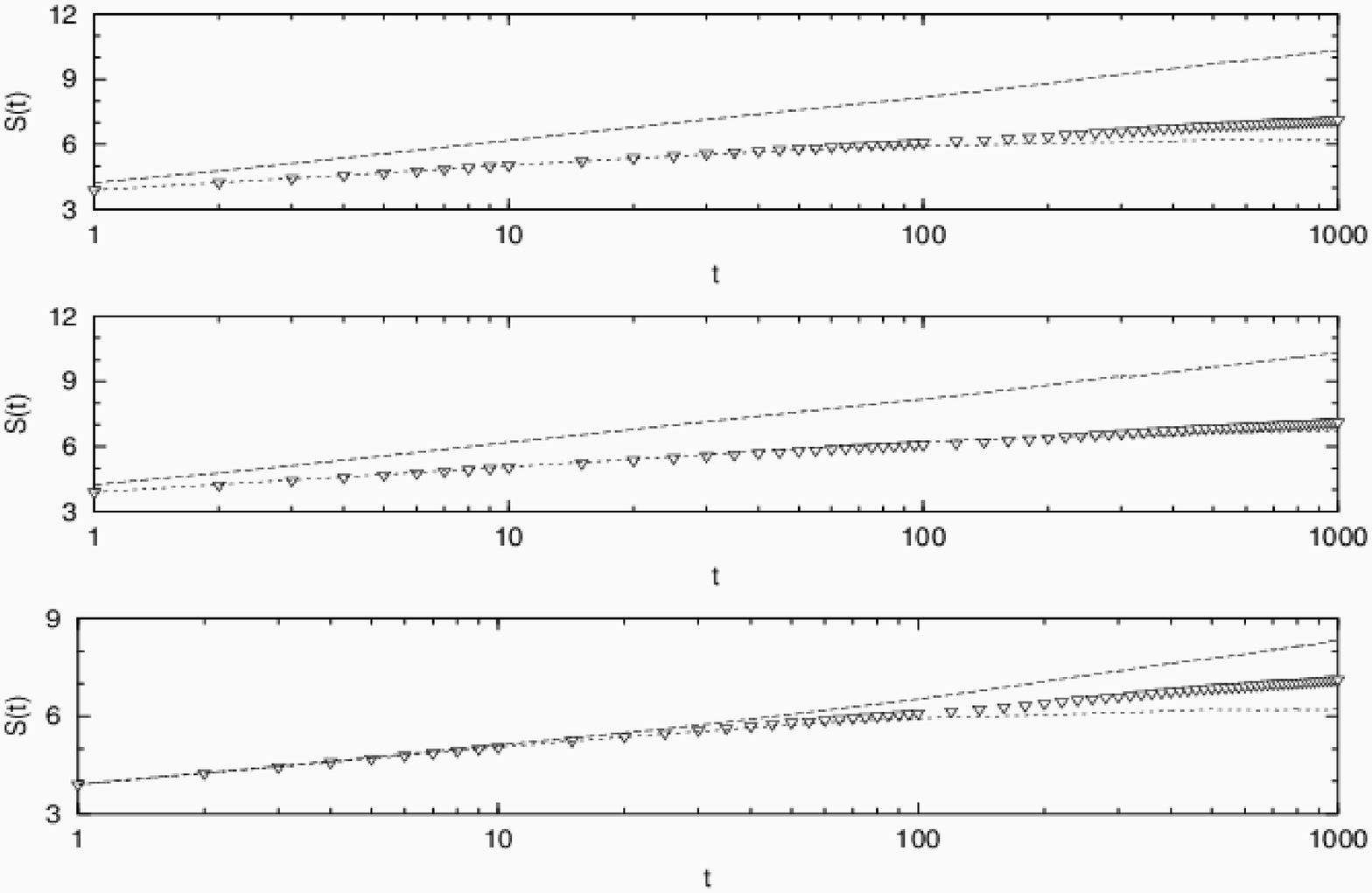}
\caption{The diffusion entropy, $S(t)$, as a function of time $t$, in a
logarithmic time scale. The triangles denote the results of the DE 
analysis
applied to the series produced only by the noise component. The dashed 
line
refers to the DE applied to the sum of noise and slow component. The 
dotted
line denotes the results of the detrending procedure. From top to bottom 
the
three frames refere $SS_{j}^{T}$, $SC_{j}^{T}$ and $SL_{j}^{T}$,
respectively. The detrending mehod used is the step smoothing procedure}
\label{figure10}
\end{figure}

\begin{figure}[tbp]
\includegraphics[angle=-90,width=5.5in] {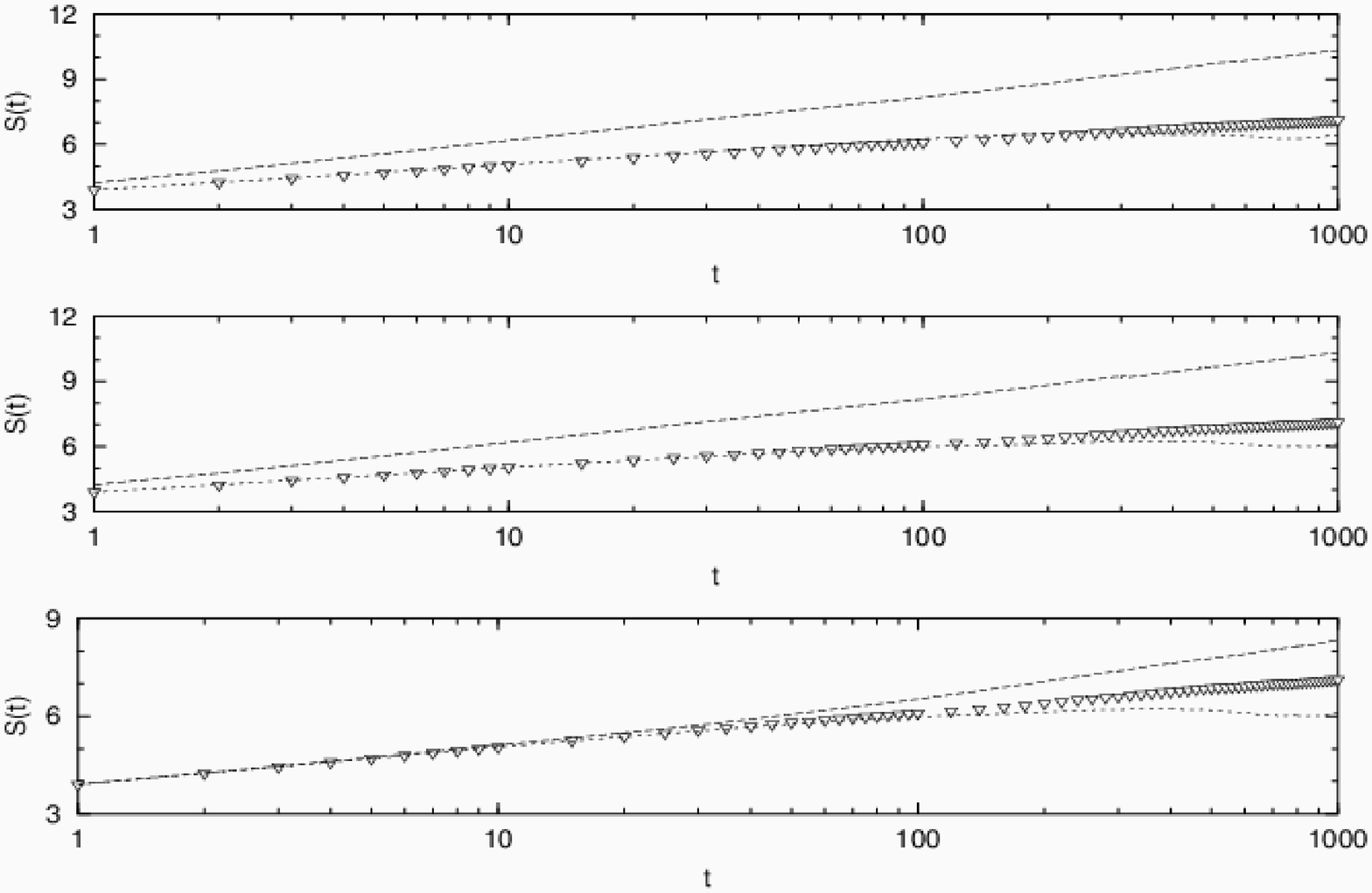}
\caption{The diffusion entropy, $S(t)$, as a function of time $t$, in a
logarithmic time scale. The triangles denote the results of the DE 
analysis
applied to the series produced only by the noise component. The dashed 
line
refers to the DE applied to the sum of noise and slow component. The 
dotted
line denotes the results of the detrending procedure. From top to bottom 
the
three frames refere $SS_{j}^{T}$, $SC_{j}^{T}$ and $SL_{j}^{T}$,
respectively. The detrending mehod used rests on the wavelet db8. }
\label{figure11}
\end{figure}

\begin{figure}[tbp]
\includegraphics[angle=-90,width=5.5in] {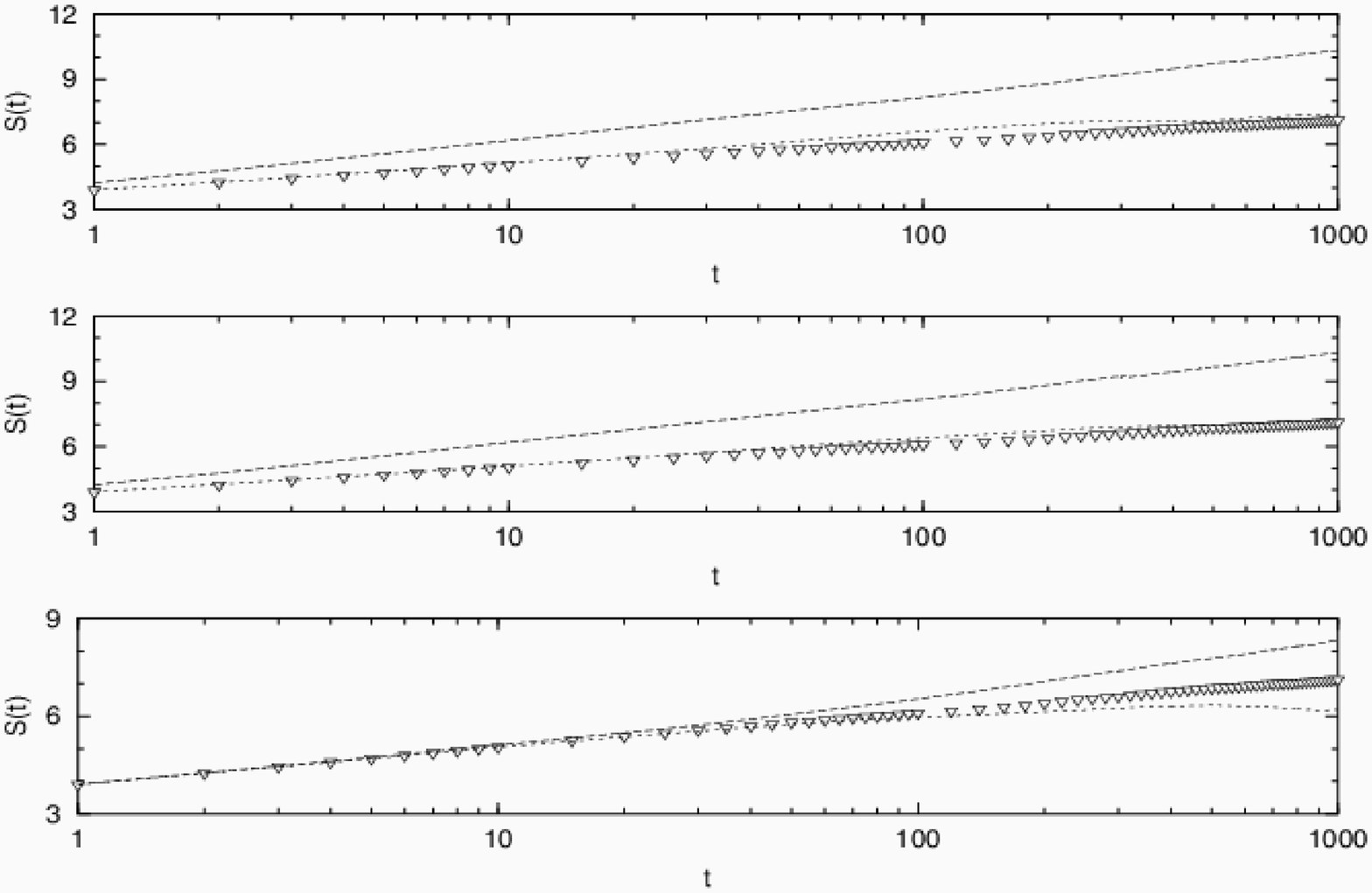}
\caption{The diffusion entropy, $S(t)$, as a function of time $t$, in a
logarithmic time scale. The triangles denote the results of the DE 
analysis
applied to the series produced only by the noise component. The dashed 
line
refers to the DE applied to the sum of noise and slow component. The 
dotted
line denotes the results of the detrending procedure. From top to bottom 
the
three frames refere $SS_{j}^{T}$, $SC_{j}^{T}$ and $SL_{j}^{T}$,
respectively. The detrending mehod used rests on the wavelet db1.}
\label{figure12}
\end{figure}

Now we apply the DE method to the time series resulting from the detrending
procedures and assess the extent we succeed in recovering the statistical
properties of the GN. We consider three time series resulting from the
addition of GN to the three slow components $SS_{j}^{T}$, $SC_{j}^{T}$ and $%
SL_{j}^{T}.$ In Fig. \ref{figure10} we apply the step smoothing detrending and in
Figs. \ref{figure11} and \ref{figure12} we use the wavelets db8 and db1, respectively. We notice that
in all three cases the detrending procedure works very well, if we ignore
the saturation taking place at long times. The remainder of this subsection
is devoted to explaining why the detrending procedure is affected by the
unwanted saturation effect.

Let us first direct our attention to the results in Fig. \ref{figure10}, the case of the
step smoothing procedure, and why the saturation effect is an artifact of
the detrending procedure. Using Eq. (\ref{practicalBj}) it can be shown that the
detrended time series (variable $\tilde{\xi}$) is obtained from the original
time series according to the following prescription 
\begin{equation}
\tilde{\xi}_{k}=\xi _{k}-\sum_{j}\left( \Pi _{k}^{j}\text{ }\frac{X_{j}}{T}%
\right) ,  \label{detrendedsignal}
\end{equation}
where $\Pi _{k}^{j}$ is the characteristic function of the $j$-th patch. Using
(\ref{slow+random}) and Eq. (\ref{twopieceszj}), it is possible to write

\begin{equation}
\tilde{\xi}_{k}=\zeta _{k}+S_{k}^{T}-\sum_{j}\left( \Pi _{k}^{j}\text{ }%
\frac{A_{j}}{T}\right) -\sum_{j}\left( \Pi _{k}^{j}\text{ }\frac{R_{j}}{T}%
\right) .  \label{detrendedsignal2}
\end{equation}
We recall that different diffusing trajectories are created using the
method of overlapping windows. According to this method, the position
occupied at time $t$ by the $m$-th trajectory, denoted by $\Gamma _{\tilde{%
\xi}}(m,t)$, is given by 
\begin{equation}
\Gamma _{\tilde{\xi}}(m,t)=\sum_{k=m}^{k=m+t-1}\tilde{\xi}_{k}.
\label{gammaxitilde}
\end{equation}
The right-hand side of Eq. (\ref{detrendedsignal2}) is the sum of four
contributions, and, correspondingly, the right-hand side of Eq. (\ref
{gammaxitilde}) can be expressed as the sum of four terms. These terms are,
with, obvious notations, $\Gamma _{\zeta }(m,t)$, $\Gamma _{S^{T}}(m,t)$, $%
\Gamma _{A}(m,t)$ and $\Gamma _{R}(m,t)$. Therefore Eq. (\ref{gammaxitilde}) can
be written as 
\begin{equation}
\Gamma _{\tilde{\xi}}(m,t)=\Gamma _{\zeta }(m,t)+\Gamma _{S^{T}}(m,t)-\Gamma
_{A}(m,t)-\Gamma _{R}(m,t).  \label{gammaxitilde2}
\end{equation}
At this point we can determine the reason for the saturation effect. In
fact, when the index $m$ denotes the beginning of a patch and $t=T$, the
function $\Gamma _{\tilde{\xi}}(m,t)$, being the sum of $\tilde{\xi}$ within
the patch, vanishes (see the definition of $R_{j}$ and $A_{j}$ in Eq. (\ref
{definitionofRjAj})). For values of $m$ denoting a position different from
the first site of a patch, the quantity $\Gamma _{\tilde{\xi}}(m,t)$ can
assume non-vanishing values. However, the above mentioned constraint
establishes an upper bound on $\Gamma _{\tilde{\xi}}(m,t)$, thereby reducing
the spreading of diffusion process. As we can see from Fig. \ref{figure10}, the
reduction of the diffusion process is already significant at times smaller
than $T$. Consequently, the step detrending procedure successfully detrends
for only a limited range of times, which we estimate to be $t<\frac{T}{3}$.

The wavelet method yields saturation effects for similar reasons. In fact
using the wavelet method we detrend the $j$-th approximation ($2^{j}\gtrsim T$
), and the $j$-th approximation is obtained through a filtering process
averaging all the components with a time scale smaller than $2^{j}$.
Therefore, this procedure is similar to the step smoothing, and the ensuing
saturation effects have the same origin.

Finally, we notice that the step smoothing is slightly more effective than
the wavelet db8 smoothing in the case of the step component $SS_{j}^{T}$,
while in the case of the continuous component $SC_{j}^{T}$ it is the other
way around. In fact, the nature of the two signals $SS_{j}^{T}$
and $SC_{j}^{T}$ is such that the step smoothing is naturally the best
``fit" for $SS_{j}^{T}$ and the db8 wavelet smoothing is naturally the best
one for $SC_{j}^{T}$. Therefore, in the case of the step smoothing acting on 
$SC_{j}^{T}$, the detrending process in not so successful and its effect
occur before saturation, becoming thus visible. The same argument applies to
the db8 wavelet smoothing acting on $SS_{j}^{T}$, thereby explaining why the
db1 wavelet smoothing does not produce excellent results when applied to
both $SS_{j}^{T}$ and $SC_{j}^{T}$ (top and middle frame of Fig. \ref{figure12}. Finally,
in the case of the $SL_{j}^{T}$ all the detrending procedures do not work
effectively, given the fact that in this case the noise intensity is much
greater than the slow-component intensity.

\section{Computer-generated Data: Effects of a Periodic Component}\label{cpu-periodiccomponent}

In this section we analyze, again with the help of computer-generated
sequences, the effects of a periodic component rather than the slow
components. This is another issue of crucial importance for the analysis of
data sets with strong annual periodicities. The effects of this kind of
external bias have been previously studied in Refs. \cite{de3,de6}. The task
of this section is, of course, that of proving that the detrending of a
periodic component is a necessary step toward the detection of the true
scaling of the underlying phenomenon with the diffusion based methods of
analysis, DE, SM, MS and DAS.

We use the same notation as that of Eq. (\ref{slow+random}), and we describe the signal
to analyze in this section as 
\begin{equation}
\xi _{j}=\Phi _{j}^{T}+\zeta _{j},  \label{genseasonal}
\end{equation}
where $\Phi _{j}^{T}$ is a periodic function, of period $T$, satisfying the
relation 
\begin{equation}
\sum_{k=j}^{k<(j+1)T}\Phi _{k}=0\quad \forall j.  \label{zeromean}
\end{equation}
Eq. (\ref{zeromean}) yields the vanishing of the mean value of function $%
\Phi _{j}$. As in the case studied in the previous Section, the presence of
the seasonal component has the effect of ``masking" the scaling properties
of the diffusion process stemming from the variable $\zeta $, by producing
additional spreading. However, due to the fact that the external bias is
periodic, the additional spreading undergoes regression to the initial
condition at times that are an integer multiple of the period $T$. This
yields an alternating sequence of increasing and decreasing spreading
phases. For this reason we refer to the diffusion effect caused by the
periodic component as the ``accordion" effect.

For a convenient explanation of the ``accordion" effect, let us study the
variable $\Gamma _{\xi }(j,t)$. As in the previous section, the quantity $t$
is the length of the window (or diffusion time) and $j$ is the initial
position of the sliding window. In the present case $\Gamma _{\xi }(j,t)$
reads 
\begin{equation}
\Gamma _{\xi }(j,t)=\sum_{k=j}^{k=j+t-1}\left( \Phi _{k}^{T}+\zeta
_{k}\right) =\Gamma _{\Phi ^{T}}(j,t)+\Gamma _{\zeta }(j,t).
\label{fisarmonica}
\end{equation}
By noticing that 
\begin{equation}
t=nT+\tau ,  \label{parteintera}
\end{equation}
with $n$ being a given integer depending on $t$ and $\tau ,$ a real number
between $0$ and $T$, and by using Eq. (\ref{zeromean}), we obtain

\begin{equation}
\Gamma _{\xi }(j,t)=\Gamma _{\Phi ^{T}}(j+nT,j+nT+\tau )+\Gamma _{\zeta
}(j,t).  \label{parteintera2}
\end{equation}
Now, with $t$ and therefore $\tau $ fixed, as the index $j$ runs along the
sequence the function $\Gamma _{\Phi ^{T}}(j+nT,j+nT+\tau )$ repeats itself
every $T$ steps and inside this time period the function assumes a maximum
and a minimum value. The difference between these two extremes is a measure
of the spreading due to the seasonal component. It is evident that if $\tau
=0$, that is, if $t$ is a multiple of the period $T$, the ``seasonal"
spreading is zero. With $\tau $ increasing, the spreading increases, but then it has, eventually, to decrease since for $\tau =T$
the initial condition must be recovered. Using Eq. (\ref{fisarmonica}) or Eq. (\ref
{parteintera2}) it is evident that if we want to know the scaling property
of the diffusion relative to the variable $\zeta $ alone, we have to look at
the diffusion relative to the variable $\xi $ only at times $t$ that are
multiples of the period $T$, since for these times, we have

\begin{equation}
\Gamma _{\xi }(j,t)=\Gamma _{\zeta }(j,t)  \label{sincrono}
\end{equation}
throughout the whole sequence. So, in principle, there would be no reason to
process the data in any way in order to retrieve the desired information on
the variable $\zeta $. We might, in fact, limit ourselves to studying the
behavior of the DE or SM at times that are a multiple of $T$. However, when
dealing with real data, the limitation on the number of data points
available and the excessive magnitude of $T$ limits the accuracy of the
observation of times that are multiples of the time period. For example if
the number of data is $13149$ and $T=365$ (the case study in paper II),
then the saturation effect, due to lack of statistics, affects the DE or SM
already at times smaller than $2T$. Therefore if we want insight into the
properties of the diffusion process of the noise component, before any
saturation takes place, we must detrend the periodic component from the time
series.

Let us proceed with the detrending in this case. Consider a time series
whose length is $NT$, $N$ being an integer. In other words, we assume the
sequence to have a length which is a multiple of the time period $T$. We
define 
\begin{equation}
\Sigma _{\xi }(j)=\sum_{m=0}^{N-1}\xi _{j+mT}=N\Phi
_{j}+\sum_{m=0}^{N-1}\zeta _{j+mT}  \label{sigmas}
\end{equation}
for all the values of $j\in \left[ 0,T\right] $, or, in a more concise
notation,

\begin{equation}
\Sigma _{\xi }(j)=N\Phi _{j}+\Sigma _{\zeta }(j)\quad j\in \left[ 0,T\right]
.  \label{sigmas2}
\end{equation}
Eq. (\ref{sigmas2}) can be used to evaluate the periodic component $\Phi
_{j}$ if  $\Sigma _{\zeta }(j)\ll N\Phi _{j}$. Using the same argument as
that adopted earlier for the detrending of the slow component. We establish
the condition

\begin{equation}
\sigma _{0}\text{ }N^{\delta }\ll N\text{ }\Phi _{j}^{T}\text{ }%
\Leftrightarrow \sigma _{0}\ll \Phi _{j}^{T}\text{ }N^{1-\delta },
\label{possibletodetrend}
\end{equation}
yielding 
\begin{equation}
\Phi _{j}^{T}\approx \frac{\Sigma _{\xi }(j)}{N}.  \label{valueofBper}
\end{equation}

We note a significant difference with respect to the case of the moving
window that we are forced to adopt in the general case. In this case the
detrended data do not show any saturation effect. In fact 
\begin{equation}
\tilde{\xi}_{j}=\zeta _{j}-\frac{1}{N}\sum_{m=0}^{N-1}\zeta _{j+mT},
\label{detrendenddataseason}
\end{equation}
does not vanish if the sum of $\tilde{\xi}_{j}$ is carried out over a patch.

\begin{figure}[tbp]
\includegraphics[angle=-90,width=5.5in] {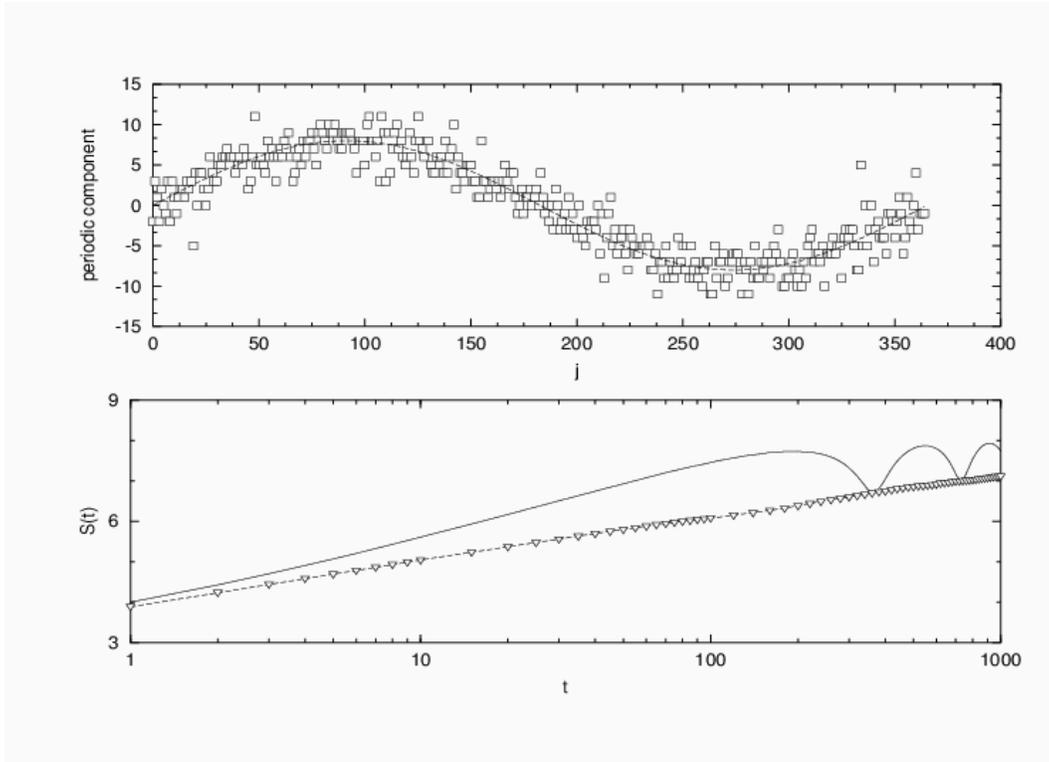}
\caption{Top frame: The periodic component as a function of the linear 
time.
The full lines denotes the real component and the squares the results of 
a
procedure based on the use of Eq.(\ref{valueofBper}). Bottom frame:
diffusion entropy $S(t)$ as a function of time $t$ in a logarithmic time
scale. The triangles denotes the result of the DE analysis applied to 
the
noise component alone. The full line illustrates the result of the DE
analysis applied to the time series stemming from the sum of noise and
periodic component. The dashed line refers to the the results of the DE
applied to the signal resulting from the detrending procedure of Section 
IV
A.}
\label{figure13}
\end{figure}

In Fig. \ref{figure13} we illustrate the results of a test done with a periodic function
and a computer-generated sequence having GN. The first frame of Fig. \ref{figure13} shows
the periodic component used and the one resulting from the detrending
procedure described above, the agreement is good. The second frame depicts
the diffusion entropy applied to the sequence with both noise and periodic
component, to the sequence with only the noise component and to the sequence
derived from the detrending procedure. The agreement between the DE applied
to the original noise and the DE applied to the detrended sequence is
impressive, and, as expected, there is no indication of a
saturation effect.

\section{Concluding Remarks}\label{fi}

We have shown that the DE method, although possessing the
remarkable property of detecting genuine scaling if it exists, cannot be
trusted in general as method for detecting the true 
scaling index. In fact, the existence of scaling makes the
diffusion entropy grow in direct proportion to the logarithm of
time. However, if the diffusion entropy grows in direct
proportion to the logarithm of time, this growth is not
necessarily an indication that the rate of this logarithmic
increase is the scaling parameter. It is possible that there is no
scaling \cite{de7}. Nevertheless, the adoption of the DE method is
useful, since it allows us to determine very quickly the range of values 
over which the scaling parameter may be located. At that stage the
determination of the correct scaling value of the scaling parameter
has to be determined using a more fine grained analysis,
based on the MS analysis and the DAS. In paper II \cite{companion} we shall see this method at work in a case of sociological interest.

MI and PG gratefully acknowledge financial support from the Army Research Office Grant DAAD 19-02-0037. PH acknoledges support from NICHD grant R03

\newpage

\end{document}